\documentclass[preprints,article,accept,moreauthors,pdftex]{Definitions/mdpi} 

\firstpage{1} 
\pubvolume{1}
\issuenum{1}
\articlenumber{0}
\pubyear{2023}
\copyrightyear{2023}
\externaleditor{{Academic Editor: Bidzina Kapanadze}}
\datereceived{27 May 2023} 
\dateaccepted{29 June 2023} 
\datepublished{} 
\hreflink{https://doi.org/} 

\Title{Observational Implications of OJ 287's Predicted 2022 Disk Impact in the Black Hole Binary Model}

\TitleCitation{Observational Implications of OJ 287's Predicted 2022 Disk Impact in the Black Hole Binary Model}

\Author{Mauri J. Valtonen
 $^{1,2,}$*\orcidA{}, Lankeswar Dey $^{3}$\orcidB{}, Achamveedu 
 Gopakumar $^{3}$\orcidC{}, Staszek Zola $^{4}$\orcidD{}, Anne L\"ahteenm\"aki $^{5,6}$, Merja~Tornikoski $^5$, Alok C. Gupta $^{7,8}$, 
Tapio Pursimo $^{9}$, Emil Knudstrup $^9$, Jose L. Gomez $^{10}$\orcidF{}, \linebreak  Rene Hudec $^{11,12,13}$\orcidG{}, Martin Jel\'{\i}nek $^{12}$, Jan \v{S}trobl $^{12}$, Andrei V. Berdyugin $^{2}$, Stefano Ciprini $^{14,15}$, \linebreak  Daniel E. Reichart $^{16}$, Vladimir~V.~Kouprianov~$^{16}$,
Katsura Matsumoto $^{17}$, Marek Drozdz $^{18}$, Markus Mugrauer $^{19}$, Alberto Sadun $^{20}$, Michal~Zejmo~$^{21}$, Aimo Sillanp\"a\"a $^2$, Harry J. Lehto $^2$, Kari Nilsson $^1$, Ryo Imazawa $^{22}$ \linebreak  and Makoto Uemura $^{23}$}

\AuthorNames{Mauri Valtonen, Lankeswar Dey, A. Gopakumar, Staszek Zola, Anne L\"ahteenm\"aki, Merja Tornikoski, Alok C. Gupta, Tapio Pursimo, Emil Knudstrup, Jose L. Gomez, Rene Hudec, Martin Jel\'{\i}nek, Jan \v{S}trobl, Andrei Berdyugin, Stefano Ciprini, Daniel E. Reichart, Vladimir V. Kouprianov, Katsura Matsumoto, Marek Drozdz, Markus Mugrauer, Alberto Sadun, Michal Zejmo, Aimo Sillanp\"a\"a, Harry J. Lehto, Kari Nilsson, Ryo Imazawa and Makoto Uemura}

\AuthorCitation{Valtonen, M.; Dey, L.; Gopakumar, A.; Zola, S.; L\"ahteenm\"aki,~A.; Tornikoski, M.; Gupta, A.C.; Pursimo, T.; Knudstrup,~E.; Gomez,~J.L.; et al.}

\address{%
$^{1}$ \quad Finnish Centre for Astronomy with ESO (FINCA), University of Turku, 20014 Turku, Finland; kani@utu.fi 
\\
$^{2}$ \quad Tuorla Observatory, Department of Physics and Astronomy, University of Turku, 20014 Turku, Finland; andber@utu.fi (A.V.B.); aimo.sillanpaa@utu.fi (A.S.); harry.lehto@utu.fi (H.J.L.)\\
$^{3}$ \quad Department of Astronomy and Astrophysics, Tata Institute of Fundamental Research, Mumbai 400005, India; lankeswar.dey@tifr.res.in (L.D.); gopu.tifr@gmail.com (A.G.)\\
$^{4}$ \quad Astronomical Observatory, Jagiellonian University, ul. Orla 171, 30-244 Krakow, Poland; szola@oa.uj.edu.pl\\
$^5$ \quad Mets\"ahovi Radio Observatory, Aalto University, Mets\"ahovintie 114, 02540 Kylm\"al\"a, Finland; anne.lahteenmaki@aalto.fi (A.L.); merja.tornikoski@aalto.fi (M.T.)\\
$^6$ \quad Department of Electronics and Nanoengineering, Aalto University, P.O. Box 15500, 
 00076 
 Aalto, 
 Finland\\
$^7$ \quad Aryabhatta Research Institute of Observational Sciences (ARIES), Manora Park, Nainital 263001, India; acgupta30@gmail.com (A.C.G.)\\
$^{8}$ \quad Key Laboratory for Research in Galaxies and Cosmology, Shanghai Astronomical Observatory, Chinese Academy of Sciences, Shanghai 200030, China\\
$^{9}$ \quad Nordic Optical Telescope, Apartado 474, E-38700 Santa Cruz de La Palma, Spain; tpursimo@not.iac.es (T.P.); emil@phys.au.dk (E.K.)\\
$^{10}$ \quad The Institute 
 of Astrophysics of Andalusia---CSIC, Glorieta de la Astronomia s/n, 18008 Granada, Spain; jlgomez@iaa.csic.es\\
$^{11}$ \quad Faculty of Electrical Engineering, Czech Technical University, 166 36 Prague, Czech Republic; rene.hudec@gmail.com\\
$^{12}$ \quad Astronomical Institute (ASU CAS), 251 65 Ond\v{r}ejov, Czech Republic; martin.jelinek@asu.cas.cz (M.J.); strobl@asu.cas.cz (J.\v{S}.)\\
$^{13}$ \quad Engelhardt Observatory, Kazan Federal University, 
Kazan 420008, Russia\\
$^{14}$ \quad National Institute 
 of Nuclear Physics (INFN), Sezione di Roma Tor Vergata, Via della Ricerca Scientifica 1, 00133 Rome, Italy; stefano.ciprini@ssdc.asi.it\\
$^{15}$ \quad ASI Space Science Data Center (SSDC), Via del Politecnico, 00133 Rome, Italy\\
$^{16}$ \quad Department of Physics and Astronomy, University 
 of North Carolina at Chapel Hill, Chapel Hill, NC 27599, USA; dan.reichart@gmail.com (D.E.R.); v.kouprianov@gmail.com (V.V.K.)\\
$^{17}$ \quad Astronomical Institute, Osaka Kyoiku University, 4-698 Asahigaoka, Kashiwara, Osaka 582-8582, Japan; katsura@cc.osaka-kyoiku.ac.jp\\
$^{18}$ \quad Mt. Suhora Observatory, Pedagogical University, ul. Podchorazych 2, 30-084 Krakow, Poland; sfdrozdz@cyf-kr.edu.pl\\
$^{19}$ \quad Astrophysical 
 Institute and University Observatory, Schillergässchen 2, D-07745 Jena, Germany; markus@astro.uni-jena.de\\
$^{20}$ \quad Department of Physics, University of Colorado, Denver, CO 80217, USA; alberto.sadun@ucdenver.edu\\
$^{21}$ \quad Kepler Institute of Astronomy, University of Zielona Gora, Lubuska 2, 65-265 Zielona Gora, Poland; michalzejmo@gmail.com\\
$^{22}$ \quad Department of Physics, Graduate School of Advanced Science and Engineering, Hiroshima University,  1-3-1 Kagamiyama, Higashi-Hiroshima, Hiroshima 739-8526, Japan; imazawa@astro.hiroshima-u.ac.jp\\
$^{23}$ \quad Hiroshima Astrophysical Science Center, Hiroshima University, 1-3-1 Kagamiyama, Higashi-Hiroshima, Hiroshima 739-8526, Japan; uemuram@hiroshima-u.ac.jp}

\corres{Correspondence: mvaltonen2001@yahoo.com}

\abstract{We present a summary of the results of the OJ~287 observational campaign, which was carried out during the 2021/2022 observational season. This season is special in the binary model because the major axis of the precessing binary happens to lie almost exactly in the plane of the accretion disc of the primary. This leads to pairs of almost identical impacts between the secondary black hole and the accretion disk in 2005 and 2022. In 2005, a special flare called ``blue flash'' was observed 35 days after the disk impact, which should have also been verifiable in 2022. We did observe a similar flash and were able to obtain more details of its properties. We describe this in the framework of expanding cloud models. In addition, we were able to identify the flare arising exactly at the time of the disc crossing from its photo-polarimetric and gamma-ray properties. This is an important identification, as it directly confirms the orbit model. Moreover, we saw a huge flare that lasted only one day. We may understand this as the lighting up of the jet of the secondary black hole when its Roche lobe is suddenly flooded by the gas from the primary disk. Therefore, this may be the first time we directly observed the secondary black hole in the OJ~287 binary system.}

\begin{document}

\section{Introduction}

\begin{figure}[H]
\centering
\includegraphics[width=0.65\textwidth]{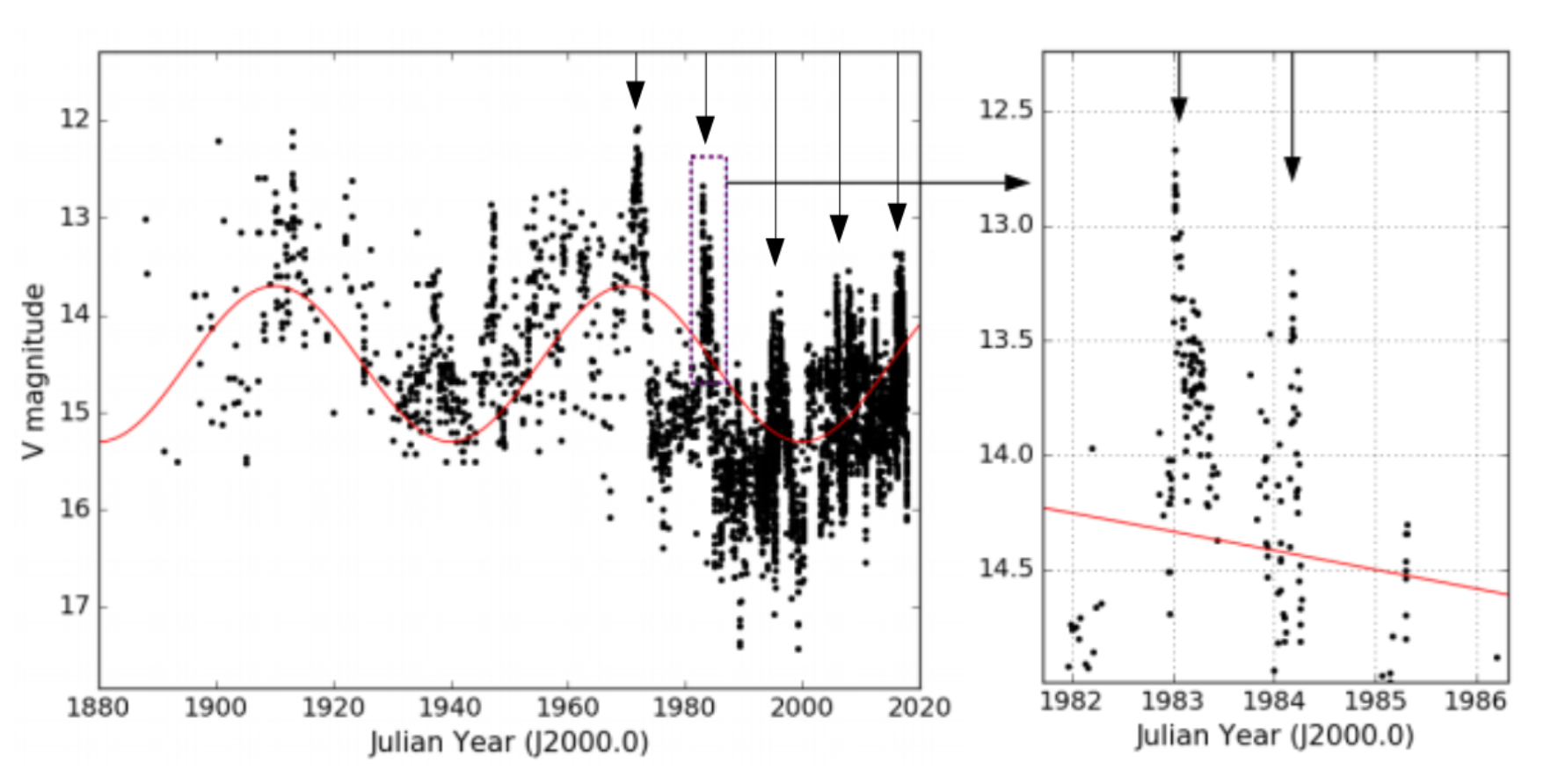}
	\caption{Optical lightcurve of OJ~287 between 1888 and 2020. The~arrows point to the 12-year orbital cycles while the line indicates the effect of the 109-year precessional cycle. Because~of planar symmetry, the~109 yr cycle appears only as half the length of the precession cycle in the light curve. The~insert from 1982 to 1986 displays the typical double peak structure. The~peaks arise from impacts on the disk, which result in the release of hot radiating plasma clouds on both sides of the disk. Therefore, for~the light curve, it does not matter whether the impact came from above or below the disk, as~seen by us, and~this results in planar symmetry with regard to the generation of optical emission. Note that the easily discernible 12-year pattern shown by arrows is lost when the major axis is close to the disk plane, as~demonstrated in \citep{val23}. This occurs during the cycle preceding the 1970s, and the arrows are not drawn further back in time. The~maxima of the precession cycle occur one 12-year cycle after the major-axis alignments, as~explained in theoretical models \citep{sun96,sun97}. The~figure first appeared in \citep{Dey18}.}
     \label{fig:lightcurve-optical}
\end{figure}

The bright blazar OJ~287 (RA: 08:54:48.87, Dec: +20:06:30.6), situated at a redshift of $z = 0.306$, has an optical light curve that goes all the way back to the year 1888. It was often unintentionally photographed due to its proximity to the ecliptic plane \citep{sil88, Hudec2013}. The~light curve exhibits unique quasi-periodic high-brightness flares (outbursts) with a period of $\sim\!12$ years \citep{sil88, val06, kid07, Dey18}.
OJ~287 also shows a long-term variation in its apparent magnitude over a period of $\sim\!55$ years. The~periods have been confirmed by a statistical study where the flare timings play no role \citep{val06} (see Figure~\ref{fig:lightcurve-optical}). The~exact values depend on the binning of the data;~for example, the 55 yr component has an uncertainty of $10\%$ arising from this source. We will mention the 109 year interval below, which is about twice the 55 yr component, after~which the flare pattern starts to approximately repeat itself.

This type of periodic behaviour was first noted by one of us (A.S.) in 1982, which lead to the successful prediction of the 1983 flare. The~first model trying explain this behaviour was given by~\citet{sil88}, who used a binary black hole (BH) model. In~the model, the secondary orbits the primary inside the accretion disk of the latter. The~model produces one accretion peak per 12-year period when the secondary goes through the pericenter of its eccentric orbit. The~model predicted the 1994 big flare~correctly.

However, with~more data from the historical light curve, it became obvious that the system is not strictly periodic. It appears that there needs to be a second flare per each 12-year cycle, and~moreover, these pairs of flares do not repeat themselves in a periodic manner. A~second, longer, period was~required.

\begin{figure}[H]
\centering
\includegraphics[width=0.65\textwidth]{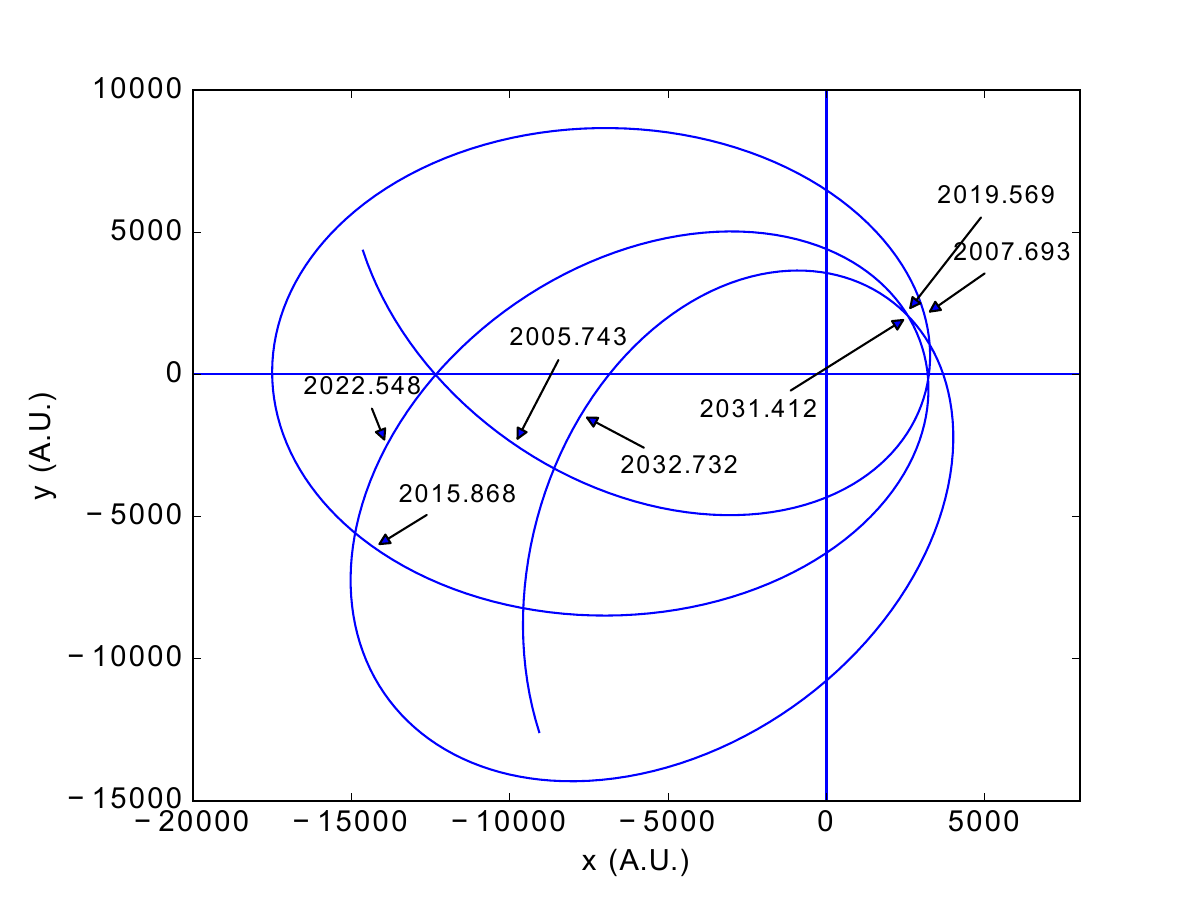}
\caption{The 
 orbit of the secondary black hole around the primary from 2004 to 2035. The~horizontal line represents the mean disk, and~is very close to the actual disk at most impacts. Arrows point to the positions of the secondary black hole at the time of the big flares. The~figure first appeared in \citep{Dey18}.\label{orbit}}
\end{figure}

In 1994, \citet{LV96} proposed a binary BH model that explains both periodicities: the~first one arising from the orbital period, the~latter from the precession period of the major axis of the binary. This leads to a pattern of flares that come mostly in pairs separated by $\sim\!1-2$ years, but~occasionally occur in triples separated by about seven years. In~this model, the~double-peaked flares occur due to the impacts of the secondary BH on the accretion disk of the primary BH twice during each orbit. The~precession of the major axis of the binary orbit has the effect of changing the time interval between the impacts. When the major axis is perpendicular to the disk plane, the~interval is at its minimum of a little more than one year. At~the other extreme, the~when the major axis lies along the disk plane, the~intervals are stretched; see Figure~\ref{orbit}.  
It is now generally agreed (see \citep{dey19}) that it is not possible to use a model that is strictly periodic to explain the historical light curve. One has to introduce a large general relativistic precession, which means that any primary mass value below $\sim\!1.5\times 10^{10}$ solar mass is strictly excluded \citep{pie98,val11b}. Dissenters like~\citet{Komossa2023} have proposed sequences that have rather more contradictions than positive correlations with the observed~lightcurve.

\textls[-20]{The model has been verified to high accuracy by a series of observing \mbox{campaigns \citep{smi85,sil85,sil96a,val96,sil96b,val06a,val08,val11b,val16}}}, as well as through a large expansion of historical data since 1994 \citep{Hudec2013}. It is important to note that the huge expansion of historical data has confirmed many flares that we did not know about previously, and~ no contradictions with the model have occurred. In~this respect, the good lightcurve coverage with upper limits seriously constrains the possibilities of alternative interpretations \citep{Dey18}.  The~latest prediction of the 2019 \textit{Eddington flare} 
 was successfully observed with the \textit{Spitzer space telescope} \citep{Laine20}, and the comparison with the 2007 light curve showed that the relative orbital times are now known with an accuracy of $\pm4$ h.

A major development in the model was the introduction of a real flexible accretion disk by~\citet{val07}. In~concrete terms, it appears as a variable disk level. From~one cycle to the next, the secondary does not find the disk at a fixed plane, but~sometimes above the mean plane and at~other times below it. The~flexibility is especially noticeable at the apocenter encounters in the eccentric orbit. The~disk level has to be calculated separately for each encounter before an orbit solution can be achieved. There is no short-cut around it: recently, Zwick and Mayer (private communication) attempted to skip this important step, but~without very promising~results.

In the present work, the~calculation of the flexible disk plays a central role, as~we are comparing two apocenter impacts: one took place in 2005 and the other occurred just recently, in 2022. The~disk calculation for the 2022 impact was not available until late in 2022, which meant that observers did not have a clear idea when to expect flares at this time. Some guesses were made, but~they were not very accurate, further confirming the importance of the actual calculation. Some observers made far-reaching conclusions based on these guesses \citep{Komossa2023}, but these have no scientific importance whatsoever once disk calculation has been~performed.

The other important step in the development of the model was the inclusion of the effect of the spin of the primary in the dynamical equations of motion \citep{val11}, as~well as bringing in the post-Newtonian corrections up to the order of 4.5 \citep{Dey18}. The~last step was verified by the \textit{Spitzer space telescope} observations \citep{Laine20}.

In addition to predicting the major impact-related flares since 1982, the~model has been able to make less precise but still interesting predictions arising from the tidal influence of the secondary on the accretion flow regarding the brightness of OJ~287 \citep{sun96,sun97,Pih13}. These predictions have also been verified \citep{val11a,val17}. The~2020 April--June flare was particularly interesting as it was predicted to be one of the brightest flares in the whole 1996--2030 prediction period, and~fully lived up to these expectations, even though it was not the very brightest of this period \citep{Komossa2020}. Similarly, the very strong predicted double flare in 2016/2017 turned out to be just as prominent as calculated in advance by~\citet{Pih13} (see~\citet{val21a}, Paper I in the following). 
 Note that the interval between the tidally induced flares depends on the orientation of the orbit in the same way as the interval between the impact flares does. 
 Therefore, the~interval from the 2016/2017 double flare to the 2020 flare is not 12 years, but~closer to 3.5 years, which is also the approximate separation of the corresponding impact~flares. 

The next opportunity after 2019 to verify or even to try to improve the orbit model will arrive during the 2031/2032 double flare. However, in~Paper I \cite{val21a}, we pointed out that there could be other ways to study the orbit, rather than relying on the direct-impact flares. The~direct-impact flare arises from a cloud of gas that has been pulled out of the disk by the impact of the secondary BH. When the cloud becomes optically thin to bremsstrahlung radiation, its radiation overwhelms the jet emission for about a week \citep{LV96}. The~pre-impact orbits were, in many ways, similar to each other in 2005 and 2022, and~we suggest making use of this rare~coincidence.

\begin{figure}[H]
  \centering
   \includegraphics[width=0.65\textwidth]{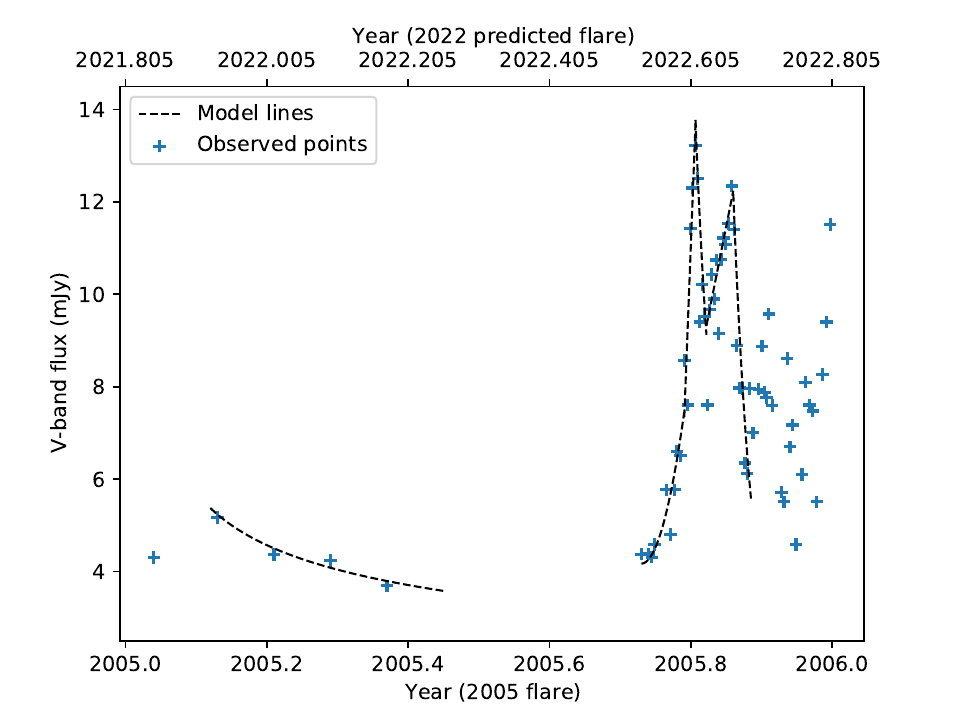}
   \caption{The optical light curve of OJ~287 during 2005 in V-band mJy units. Plot shows the monthly averaged data points for the early part of 2005, while the latter part has 3-day averaged data points. The~dashed curve provides a general outline of the brightness evolution in 2005. If, and~only if, the~disk levels and the impact distances were the same in 2005 and in 2022, one can draw an upper x-axis for the 2022 time scale. The~big flare during 2022 was not expected to be visible from any Earth-based facility. This conclusion can be reached by using the graph or by any other other reasonable criterion. The~figure first appeared in~\citet{val21a}.}
   \label{fig:2005_lc_fit}
\end{figure}

In Paper I \cite{val21a}, we showed an averaged version of the 2005 light curve, and~drew up  the scale for 2022, which could be used as a sort of prediction (see Figure~\ref{fig:2005_lc_fit}). The~light curve has at least four separate parts according to the present understanding. Throughout the whole of the year 2005, there was an underlying synchrotron emission from the primary jet. At around 2005.3, there was a brief episode of synchrotron emission from the disk, which is superposed on the jet emission. This is not clear in the monthly averaged light curve and~only becomes prominent, equal to the 2005.8 flare, when viewed in the ``BVRI light curve''. In~the latter part of 2005, the jet emission is at a high level, and~superposed on it are two peaks (outlined by the dashed line):
the bremsstrahlung peak at 2005.8, from a plasma cloud that was seen earlier (2005.3) in synchrotron emission, and~the peak at 2005.85, which is synchrotron radiation associated with the same plasma~cloud.

It is quite clear that one could not test the orbit model on this occasion, a~fact that was stressed, among~other places, in~a 2022 arXiv preprint of \citet{val23}. In~Figure~\ref{fig:2005_lc_fit}, the big flare lies in the summer period, when it is not possible to observe OJ~287 by ground-based optical telescopes due its small solar elongation. Efforts were made to find a telescope far from the Earth with a different vantage point than ground-based telescopes have. However, it was difficult to justify the use of these telescopes for our purposes, since this would have diverted them from their stated goals, and~observing OJ287 in-between their goal observations might have even jeopardised their primary mission. 
Thus, our inability to observe a big flare in 2022 is not a failure of the standard model, in~contrast to statements by \citet{Komossa2023}. There is no indication that the orbit model of~\citet{Dey18} would be in need of any upgrades or improvements. Other ideas regarding OJ~287 have been discussed in~\citet{dey19}.

Note that a prediction of the time of the big flare, or~any other flare related to the orbit, requires knowledge of both the bullet (the secondary BH) and the target (the disk). Knowledge of the disk came too late (towards the end of 2022) to enable us to make any predictions about when the bullet hit the target. Of~course, it was necessary to make conjectures about the impact event to~give observers some idea of when to carry out their observations. Some authors, particularly~\citet{Komossa2023}, seem to have confused these conjectures with real astrophysical models. We will come back to the nature of these conjectures below, since they have attracted undeserved attention in the recent~literature. 

The binary black hole parameters of the OJ~287 system, 
determined using post-Newtonian (PN) equations of motion, are: primary black hole mass $m_1 = 1.835 \times 10^{10} M_{\odot}$, secondary BH mass $m_2 = 1.5 \times 10^8 M_{\odot}$, primary BH Kerr parameter $\chi_1 = 0.38$, orbital eccentricity $e = 0.657$, and~orbital period (redshifted) $P = 12.06$ years~\cite{Dey18}.

The accuracy of the primary mass is $\pm 0.3\%$. Based on astrophysical knowledge of the system, the~mass cannot be very different from the value given above \citep{val12,val23a}. Of~course, it is possible that part of the mass measured by the model does not belong to the BH but to particles it has attracted around itself \citep{gon99}. However, it might be difficult for the BH model to achieve success in explaining the variations in the radio jet in the sky if~the attracted mass is significant, since the radio jet calculation relies on simulating the inner edge of the accretion disk \citep{Dey21}.

In the present campaign, we concentrate on the light curve in the early parts of 2005 and 2022. This is because of the secondary impacts on the disk at these times. 
 The~impact distances, and~even the impact angles are nearly the same, close to 45 degrees relative to the disk plane. However, the impact directions are different: in~2005, the impact came from outside, while in~2022, it came from inside, as seen from the primary BH (Figure \ref{orbit}). Thus, some differences may be expected, even though the basic signals arising from this impact should be similar. One of the differences is the delay between the impact and the time of the bremstrahlung flare: it is quite sensitive to the actual impact distance, and the relative disk levels at the time of impact \citep{Dey18}. 
 Therefore, the disk level in 2022 enters, in a complicated way, into the estimate of the epoch of the big bremstrahlung~flare.

We first review the disk simulations during the two impacts in 2005 and 2022. In~this way, we establish a better comparison scale than the upper axis in Figure~\ref{fig:2005_lc_fit}. We should use the disk-crossing times as reference points on the time axis rather than the midplane-crossing times. Then, we look at the two light curves at the times of the expected flares, and,~after deducting the base level flux, compare the R-band light curves of flares themselves. Next, we check if this split between the flare and the base flux agrees with the flare-to-base ratio from V/X observations by the \textit{Swift telescope}, where V is the optical V-band flux and X the X-ray flux. Here, the term ``split'' refers to the way in which we expect two different components to add up to the measured flux. This is always an~assumption.

Further, we compare the 2022 R-band and optical polarisation light curves with the model of~\citet{van71}, and~associate the flare with a definite stage in the impact simulation of~\citet{iva98}. We next determine the spectral index of the flare component from the 2005 spectral index observations and derive the 2022 spectral index light curve, assuming that the flare components have the same spectral index at both impacts. The~spectral index itself is measured in 2022, but~it is interesting to compare the measurements with what is should be if~the 2022 flux is a composite of the underlying jet emission and the added flare component. We call this flare a ``blue flash'' from the spectral region where it is primarily observed. Then, we note two other flares, which were not expected: one coinciding with exact time of impact on the disk (``Gamma-ray flare'') and the other coinciding with the lighting up of the secondary jet (``Roche-lobe flare'').

\section{Disk Impact~Simulations} 

\begin{figure}
\centering
\includegraphics[width=0.6\textwidth]{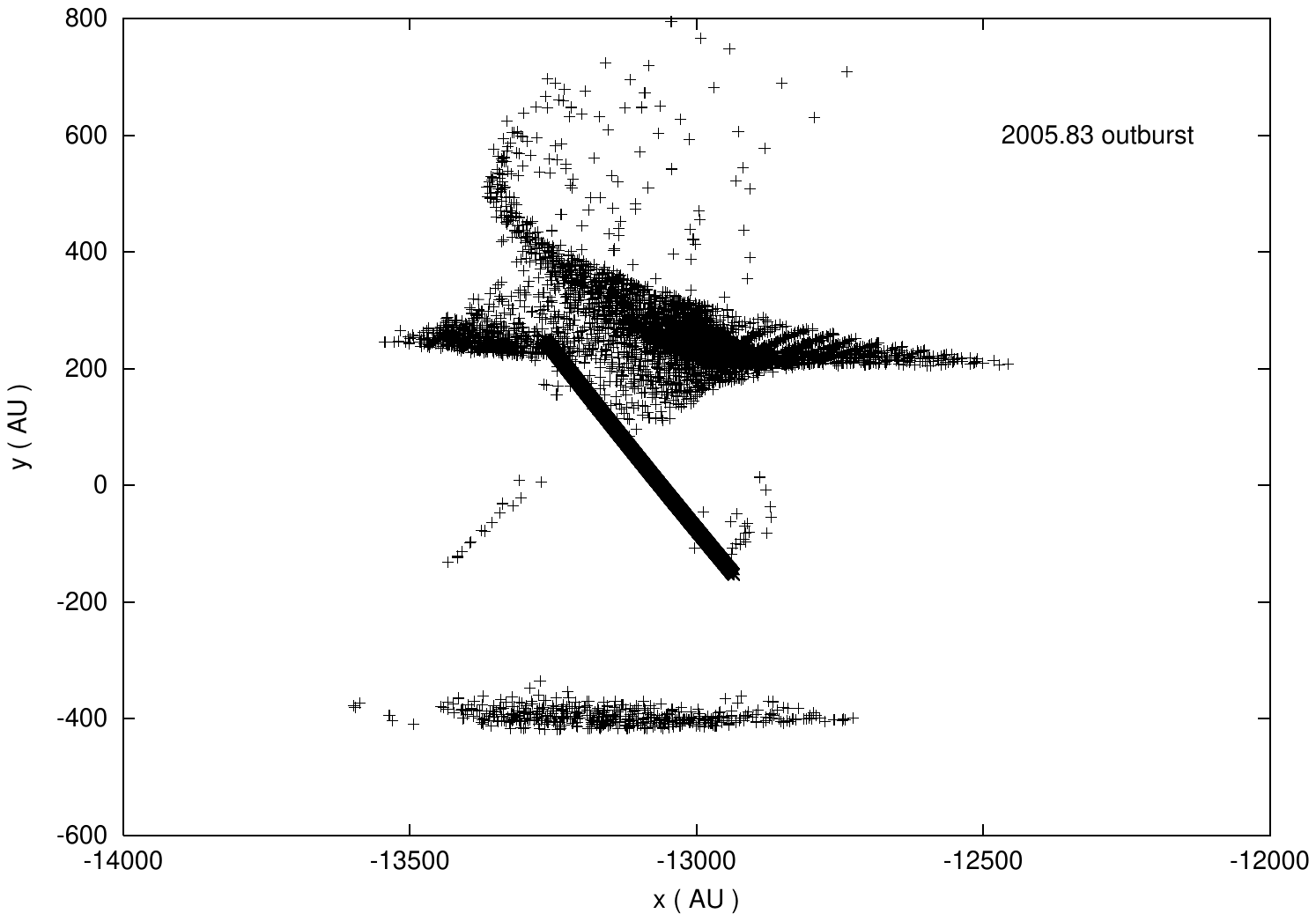}\\
\includegraphics[width=0.6\textwidth]{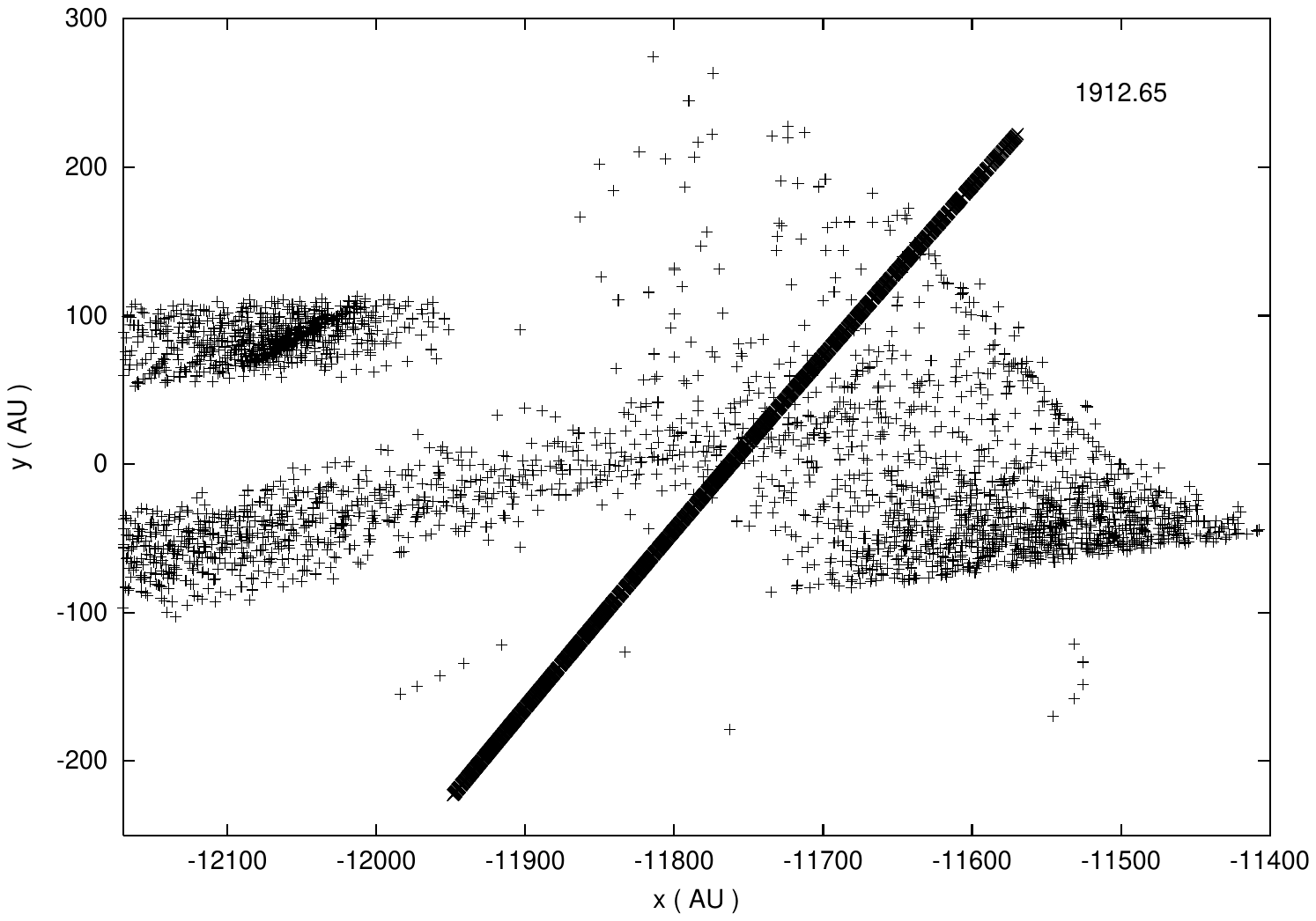}
\caption{Simulation 
 of the response of the accretion disk to the crossing of the secondary BH through the disk in 2005 (upper diagram) and 2022 (lower diagram). The~disk is represented by particles, and~this is an edge-on view of the particles that reside in the neighborhood of the impact point at the impact time. The~BH orbit is represented by a line that goes from the upper left to lower right in 2005 and from upper right to lower left in 2022. These are overlays of many snapshots, and~the BH has moved between each~snapshot.}
\label{disk05_22}
\end{figure}

The binary orbit has been solved several times with increasing accuracy \citep{LV96,val07,val10,Dey18}. In~addition to the position of the BH, we need to know the position of the disk in~order to calculate various impact-related flares. This allows for us to better line up the 2005 and 2022 light curves for comparison. We use the simulation archives of \citet{val07} and display the disk profiles in 2005 and 2022 in  Figure~\ref{disk05_22}. Actually, the~latter figure was calculated for the 1913 disk impact, but~because of the 109 yr periodicity, it applies almost perfectly for the 2022 case, except~for a 600 AU lateral shift. The~shift represents about  $5\%$ change in the radial distance, and~ introduces an uncertainty of the same relative amount in dynamical quantities such as the disk level. This is below the accuracy with which the disk levels are determined, as~we will see~below.

The disk simulations initially used a particle--hydrodynamic code \citep{sil88}, but~it was soon determined that, for the perturbing orbit of high inclination, a pure particle code is sufficient and,~being much faster, allows for high particle numbers and better statistics \citep{sun97}. Therefore, a pure particle code was used in the 2006 simulations \citep{val07}.

The simulations were started either in the 1880s or in 1940s from a disk of particles in the mean plane. After~several 12-year orbital periods, particles were scattered in the disk to correspond to the model thickness, and~the simulations were stopped when the thickness exceeded the expected thickness in the standard $\alpha$-disk model. Therefore, the simulations did not reliably represent the disk in 2022, and~the disk at that date was essentially unknown, even though the paper provided some mean values from the scattered~data.

The parameter values for the binary orbit were essentially the same as in the latest orbit model of \citet{Dey18}. Therefore, it was not necessary to redo the disk simulations at a later time. The~only free parameter is the starting date when the disk lies at its mean plane. Some experimentation showed that this is not an important parameter. The~dates we will discuss here correspond to the 1913 and 2005 impacts, both of which are far from the starting time, so the disk should have relaxed on both~occasions. 

The disk profiles near the impact site were calculated for these two dates, as~well as for many other impacts. The~disk at the impact site was represented by about 10,000 particles in the 1913 simulation and a million particles in the 2005 simulation. The~disk levels were calculated by averaging the y-coordinate for all the particles within certain radius of the site. Thus, the values are accurate to about $\sim\!5$ AU. Note that the impact site represents only a small part of the disk so that a uniformly filled disk would require particle numbers that were orders of magnitude greater. Since the particles were non-interacting, it was possible to isolate a region in the disk to obtain better~statistics.

What is important here is the level of the accretion disk, since it affects the time of the disk impact. For~the 2005 disk, the~densest part of the disk is at $y=230\pm20$ AU above the mean disk level ($y=0$). At~2022, the~disk is at $y=10\pm20$ AU. The~difference of 220 AU in the relative disk levels has an uncertainty that corresponds to the $\sim\pm3$ days uncertainty in the BH travel time between the two disk levels. The~difference agrees with \citet{val07}, who found a difference of 224 AU between the 1913 and 2005 disk levels by averaging  all disk particles near the impact site. We also note that the accuracy of the relative timing between the 2007 and 2019 disk impacts was only 4 h, because~there was no contribution from the disk level difference at that time, and~ because the crossing speed was higher. In~the present case, we expect to position the 2005 and 2022 light curves relative to each other with a $\sim\pm1$ day accuracy at best, but~the $\sim\pm3$ day accuracy may be more realistic when considering the disk-level~uncertainties.

We may view this as an astrophysical process where a bullet (BH) hits a target (disk). In~the case of 2005 and 2022, the~bullet hits a target that is practically the same in both cases. The~difference in the impact distances is of the second order, and~the way the disk parameters change over this distance range is totally negligible \citep{val19}.
For the bullet, the~important property is its momentum. Even the two components of the momentum in~2005 and 2022 are practically the same, except~for the difference in the sign of the x-component. That should not matter since the energy imparted by the bullet is independent of the sign of the momentum component. Therefore, astrophysically, the argument for the similarity of the 2005 and 2022 pair of flares is exactly the same as for the 2007 and 2019~pair.

So why did we not obtain the $\pm4$ h accuracy this time? There are two reasons: the disk-level difference was 224 AU from exact calculation, and~from the figures, we estimated this to be 220 AU. The~exact calculation refers to an annular region around the impact site, barring the very central area shown in Figure~\ref{disk05_22}, while, from the same figure, we can determine when the secondary BH crossed the highest-density region. These two independent determinations indicate that the uncertainty is at least 5 AU, and~5 AU corresponds to roughly 12 h in BH travel~time.

The second reason is that the uncertainty is related to speed: higher speed leads to a greater accuracy. From~Kepler’s second law, the transverse speed in 2022 was four times smaller than in 2019. Therefore, the $\pm4$ h accuracy degrades to $\pm16$ h accuracy. Thus, altogether, we cannot obtain a better accuracy than about 1 day for the relative timing between the 2005 and 2022 light curve events.
As we will see below, observations are consistent with the theory within one~day.


Before this calculation was recovered from the archives, we considered the possibility that the disk in 2022 could be as much as 500 AU below the mean level, which would move the big flare out of the summer ``gap'' and delay it until October. This turned out to be unrealistic, but~meant that the~observing campaign was extended. It was useful to confirm that the non-detection of the big flare is in full agreement with the theoretical calculation, as well as with earlier predictions \citep{val07,Dey18}. 

Why should we consider a disk that is so far from the mean level? In the disk simulations, we have never found such a large deviation from the mean. The~idea was not based on our binary model, but~on a certain toy model. The~model assumes that the impact of the secondary on the disk sets the phase of the brightness oscillations in the primary jet. By~correlating the observed brightness oscillations, we found that the impacts in 2005 and 2022 should be 17.0 yr apart, not 16.83 yr apart, as when the disk levels are the same. This leads to the above-mentioned disk~level.

The weakness of this toy model is the fact that it is unrelated to the standard binary model, and~that there is no evidence from previous light curve studies that it might actually work. There is also the theoretical weakness that it requires instant communication between the two black holes; otherwise, the primary black hole would not know when to choose a particular phase for the oscillations. If~we were to use the sound speed in the corona of the accretion disk as the signal speed to communicate between the two BHs, the~big flare would take place in July 2022 in the framework of this toy model. Thus, the failure to see a big flare in October 2022 tells us that instant communication is not possible, or~the toy model is not viable for other~reasons.

Some authors claim that the failure to see the October flare is a “missing flare” problem, which, in their view, leads to the need to re-evaluate all previous modelling \mbox{work on OJ287 \citep{Komossa2023}.} They also view the accurate model as a “post-factum” correction to the toy model since the accurate result was presented later than the toy model. However, the~order of presentation of the models should not matter since the accurate result was based on calculations carried out in 2006, at~a time when one could not possibly have had any information on the behaviour of OJ287 in~2022.

Therefore, we could say that we had three toy models prior to the actual model calculation: the first one assumed that the 2005 and 2022 disk levels were the same \citep{Dey18}, the~second one  additionally assumed that the impact distances were the same \citep{val21a}, and~the third one assumed that the brightness oscillations in the main jet were coordinated through instant communication with the secondary black hole (\citep{val23},the arXiv preprint mentioned above). 
The toy models were presented to help observers to coordinate their observations during the campaign, and~they were not meant to have any further significance after the actual model was calculated, in~contrast to claims by~\citet{Komossa2023}.

\section{Light Curves in 2005 and~2022} 

\begin{figure}
\includegraphics[width=0.65\textwidth]{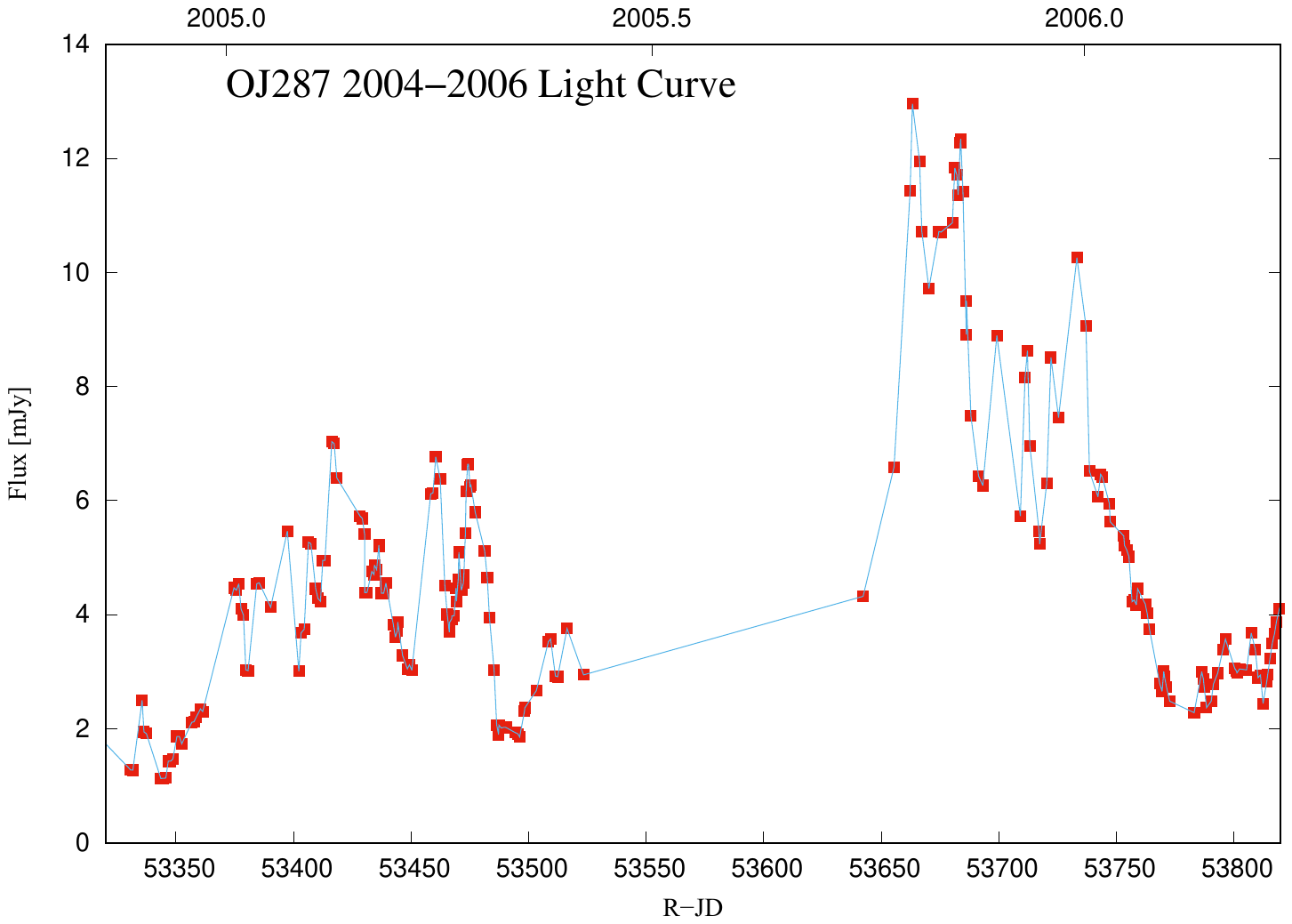}\\
\centering
\includegraphics[width=0.65\textwidth]{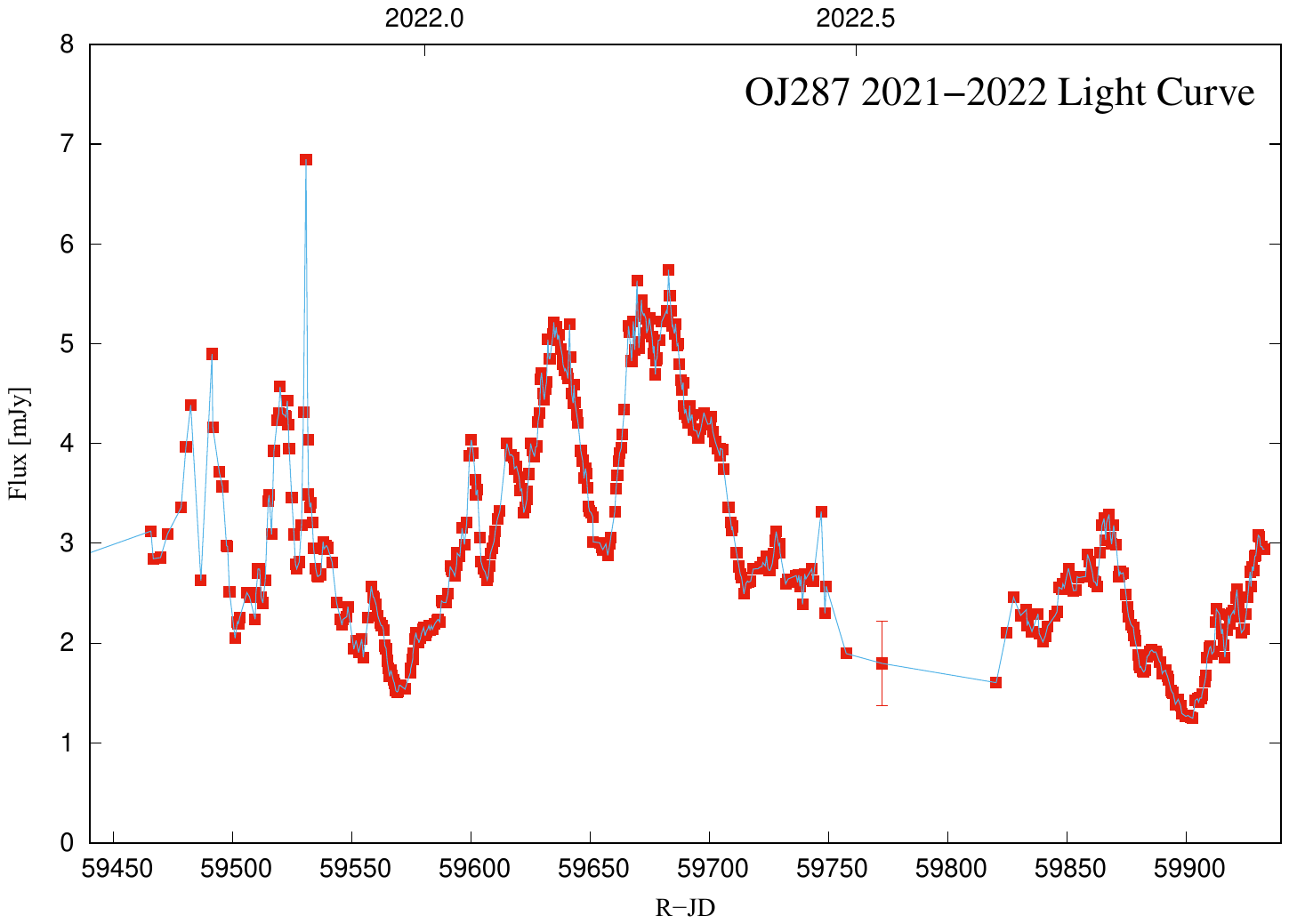}
\caption{The 
 R-band light curves of OJ287 in the 2004/05  (top panel) and 2021/22 (bottom panel) observing seasons. Since the disk level difference between 2005 and 2022 impacts is 220 AU, which corresponds to 20 days in secondary BH travel time, the~zero points for the overlay are separated by the midplane impact difference of 6140 days plus the travel time difference of 20 days, which makes the total separation  6160 days. Thus, JD 53475 in the upper figure corresponds to JD 59635 in the lower figure. In~the following, we describe the baseline as a straight line through two of the low points in both graphs (JD 53450 and 53490 in 2005; JD 59570 and 59655 in 2022). This is a simplifying assumption that allows us to study the split between the base and the flare components during those time~intervals.}
\label{lc05_22}
\end{figure}

The two light curves of~2005 and 2022 are shown in Figure~\ref{lc05_22}, where we display the data in the R-band. The~telescopes used in this work are listed in Table~\ref{telescopes}.
\unskip
\begin{table}
\centering
\caption{The telescopes and systems used in the 2021/2022 OJ~287 observing~campaign. \label{telescopes}}
\setlength{\tabcolsep}{1cm}
\begin{tabularx}{0.5\textwidth}{cc}
\toprule
\textbf{Telescope} & \textbf{Measurement}\\
\midrule
Atlas & R-band\\
Skynet & R-band\\
Krakow & R-band\\
Osaka & R-band\\
Mt. Suhora & R-band\\
Jena & R-band\\
NOT & R-band\\
Liverpool-La Palma & polarisation\\
Hiroshima & polarisation\\
Turku-Hawaii & polarisation\\
Ondrejov & BVRI spectrum\\
NASA-Swift & UV and X-rays\\
NASA-Fermi & gamma-rays\\
Mets\"ahovi & radio\\
\bottomrule
\end{tabularx}
\end{table}

Looking only at the single band data, there is nothing remarkable about these light curves. They represent mid-level activity in OJ287, where the typical properties are a rather steep spectrum from blue to infrared (BVRI), and~a degree of polarization that goes up during the flare. It is only after we study the time evolution over the whole electromagnetic band that we realize that there is some extremely rare  OJ287 behavior during these two campaign~periods.

Especially interesting is the ``BVRI light curve'', as~we will learn below. In~2005, it had two large flares: one coincided with the R-band flare in October, but~an equally large flare was seen in the spring of 2005. In~2022, we saw only the large spring flare. A~large ``BVRI flare'' may have also occurred in the summer of 2022, but OJ287 was not observable by~us at this date.

The overlay in Figure~\ref{fig:2005_lc_fit} was performed by sliding the top scale forward by 6140 days with respect to the lower scale. This is the difference in the midplane impact times in 2005 and 2022. However,~the 2022 disk level was lower than in 2005. Therefore, we have to add the travel time for the secondary BH between the two levels of 20 days and obtain a difference of 6160 days. For~example, the~recent JD 59635 flare is the counterpart of the JD 53475 flare in 2005. The~accuracy of this line-up was discussed~above.

\begin{figure}[H]
\centering
\includegraphics[width=0.55\textwidth]{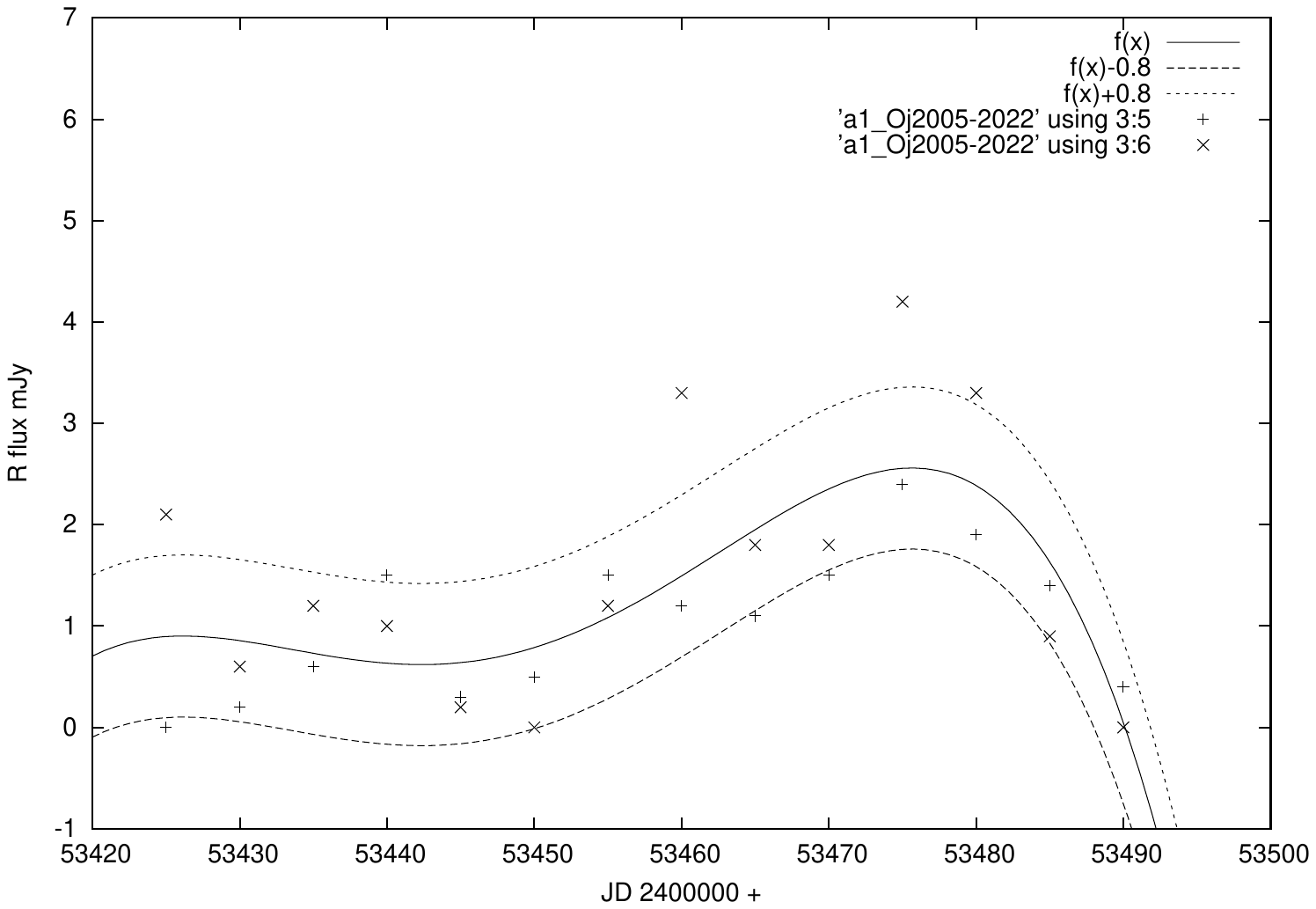}
\caption{An overlay 
 of the flare components in 2005 (from  JD 53420 to 53500, $x$ -sign) and in 2022 (from JD 59580 to 59660, $+$ -sign). The~central line is the best polynomial fit through all the points, while the two other lines show a one-sigma deviation from the fit. The~baseline is taken to be a straight line through two of the low points in both graphs, as~explained in the caption of Figure~\ref{lc05_22}. \label{base}}
\end{figure}

We may note that there is a general increase in flux around the JD 59635 flare while the general flux level was decreasing at the time of the JD 53475 flare. If~we deduct the base level in both cases using a linear rise or decline in the base level, we obtain the comparison shown in Figure~\ref{base}. The~details are explained in the figure~caption.

We note that the two flares are not significantly different from each other, considering that the background fluctuations at the level of 0.8 mJy are expected, even after the general trend is deducted~\cite{val11a}.

\section{Optical vs. X-Ray Flare in~2022}

 We may study the split between the flare and base level components by comparing the optical light curve with the X-ray light curve in 2022. For~2005, there was only a single X-ray measurement at the time, consistent with the X-rays not taking part in the flare \citep{val23}. At~the $\textit{general relativity centenary}$ flare in 2015, the~X-ray light curve did not show any unusual activity. Therefore, we may consider the X-rays as a proxy of the background emission, and~take the V/X ratio from $\textit{Swift}$ observations to represent the flare/base ratio. Here, V is the V-band flux and X is the X-ray~flux.
 
 This, of~course, is very much a hypothesis that requires experimental testing as well as theoretical understanding. This statement would be true if the main flux in optical and X-rays was formed by a single mechanism, or, at~least, by~the same radiating particles. For~example, electrons radiate by the synchrotron mechanism in the optical range and due to the inverse Compton scattering of their own synchrotron photons, they emit in the X-rays as well. Whether a reasonable model can be built that explains the V vs. X correlation seen at least occasionally in OJ287 \citep{Komossa2022} remains to be~seen.

\begin{figure}[H]
\centering
\includegraphics[width=0.65\textwidth]{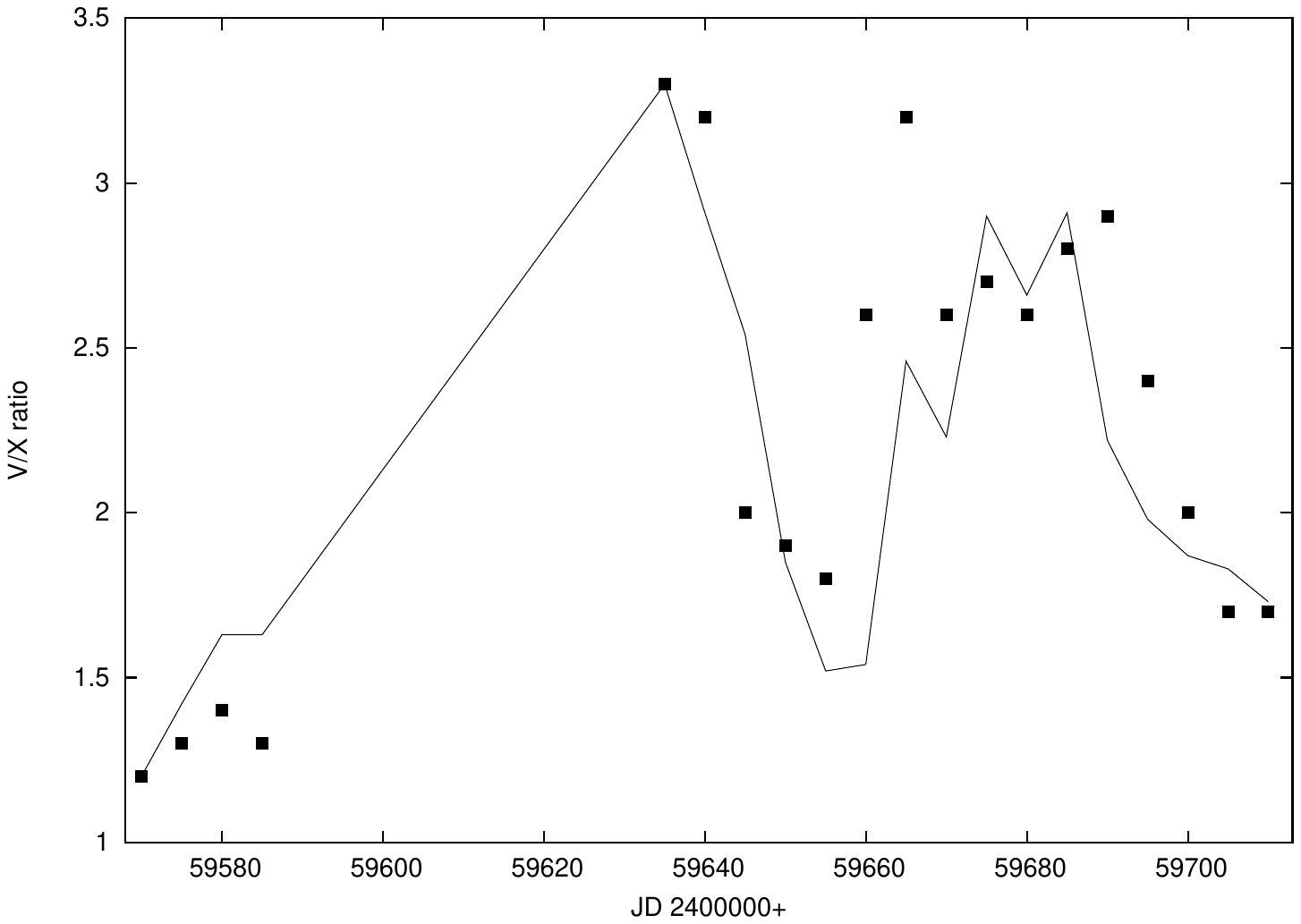}
\caption{The V/X ratio, 
where V is the observed V-band flux and X the observed X-ray flux (line) compared with the R-band flare flux over the base level flux (points). The~V/X is in arbitrary units (see~\cite{val23} for the scale; the~errors are typically 0.3 units). The~flare/base flux ratio is raised by a constant amount for easier comparison. The~fair agreement between the line and the points demonstrates that the X-ray flux is a reasonable proxy for the base flux, and~that using the straight-line approximation for the base flux evolution does not cause excessive scatter. Note that there was a long gap in \textit{Swift} observations (a long straight line on the left hand half of the figure.) \label{VX-plot2}}
\end{figure}

 The result is shown in Figure~\ref{VX-plot2}. Generally speaking, the~two ratios follow each other, which is an indication that the split used in Figure~\ref{base} is reasonable. It also shows that the flare is primarily an optical~flare.

This suggests, but~definitely does not prove, that X-rays are a good proxy to base level optical emission. The~flare itself does not appear to significantly contribute to the X-rays. This means that the spectral power-law has a break between optical and X-ray regions, which is not surprising. Due to the ageing of the relativistic electron population, the~high-end electrons vanish faster than the low-end electrons on the energy scale, and~a spectral break develops in synchrotron radiation. There is also the synchrotron self-absorption limit of optical transparency, which means that we do not expect to see any radio flux from the flare \citep{pac70}. Indeed, this is what is observed \citep{val23}. Therefore, we call the flare a ``blue flash'', since it is prominent in the optical blue~region.

\section{Spectral Index in 2005 and~2022}

One of the key observational properties is the spectral index. In~a low state, the BVRI spectral index in OJ~287 is in the range $\beta$ = 1.35--1.5 \citep{zheng2008}. The~spectral index is defined as flux $\sim$ ${\nu}^{-\beta}$, where $\nu$ is the frequency. We see that the spectrum becomes unusually flat during the JD 59635~flare. 

\begin{figure}[H]
\centering
\includegraphics[width=0.6\textwidth]{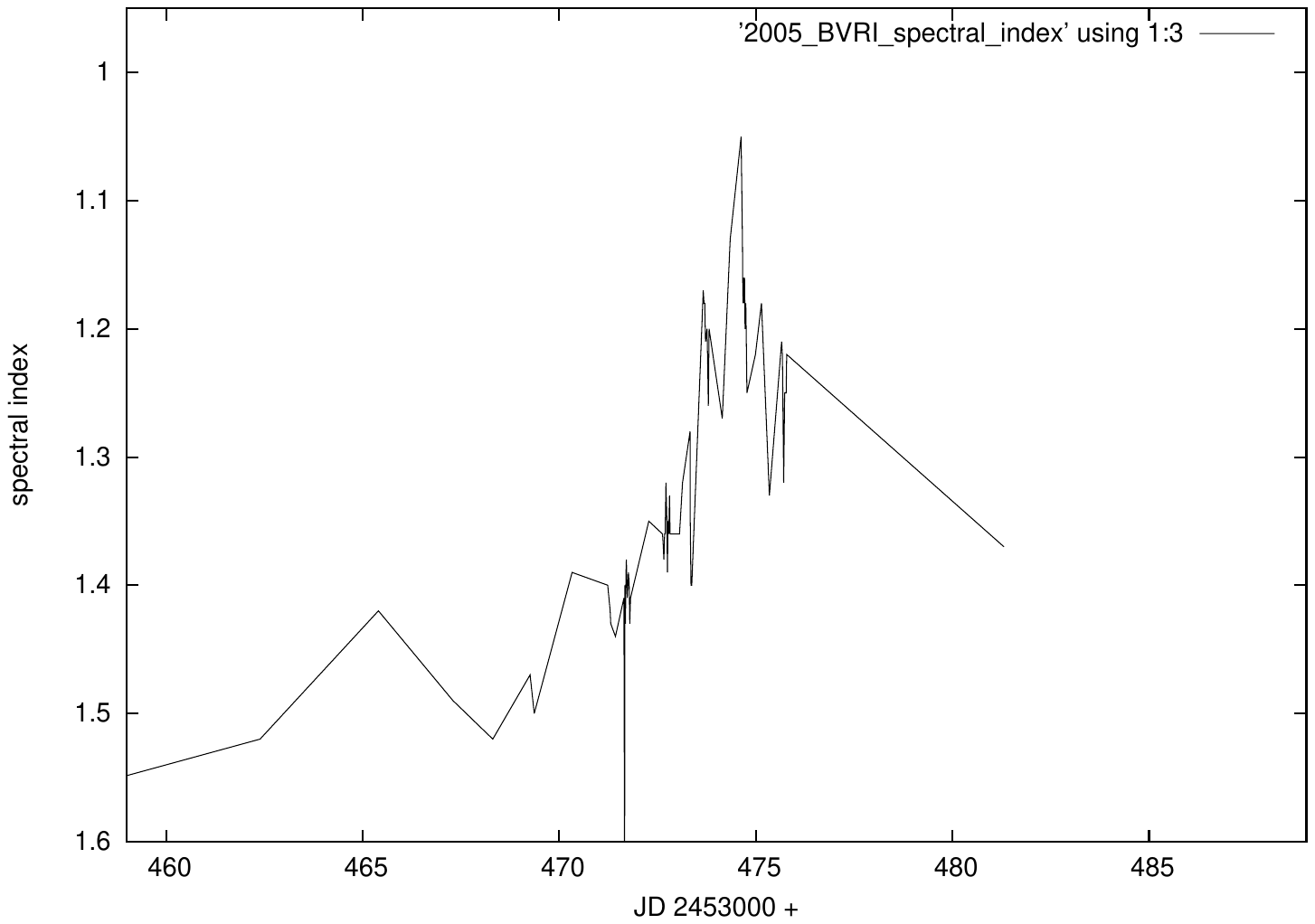}
\caption{Spectral index observations of OJ~287 in the spring of 2005 \citep{cip08}. Using the base-level spectral index $\beta=1.5$, we derive the spectral index of the flare component $\beta=0.7$.\label{sp_2005}}
\end{figure}

The JD 53475 flare stands out prominently in the spectral index light curve in the spring of 2005 (Figure \ref{sp_2005}). In~this way, it is quite different from the R-band light curve, where we find several other flares in the same brightness~range.

Assuming that the JD 53475 flare is a separate event, on top of the base level, and~using the flare vs. base split derived above, we calculate the spectral index of $\beta=0.7$ for the flare component. In~the calculation, we assume that $\beta=1.5$ for the base component \citep{cip08}.

\begin{figure}
\centering
\includegraphics[width=0.55\textwidth]{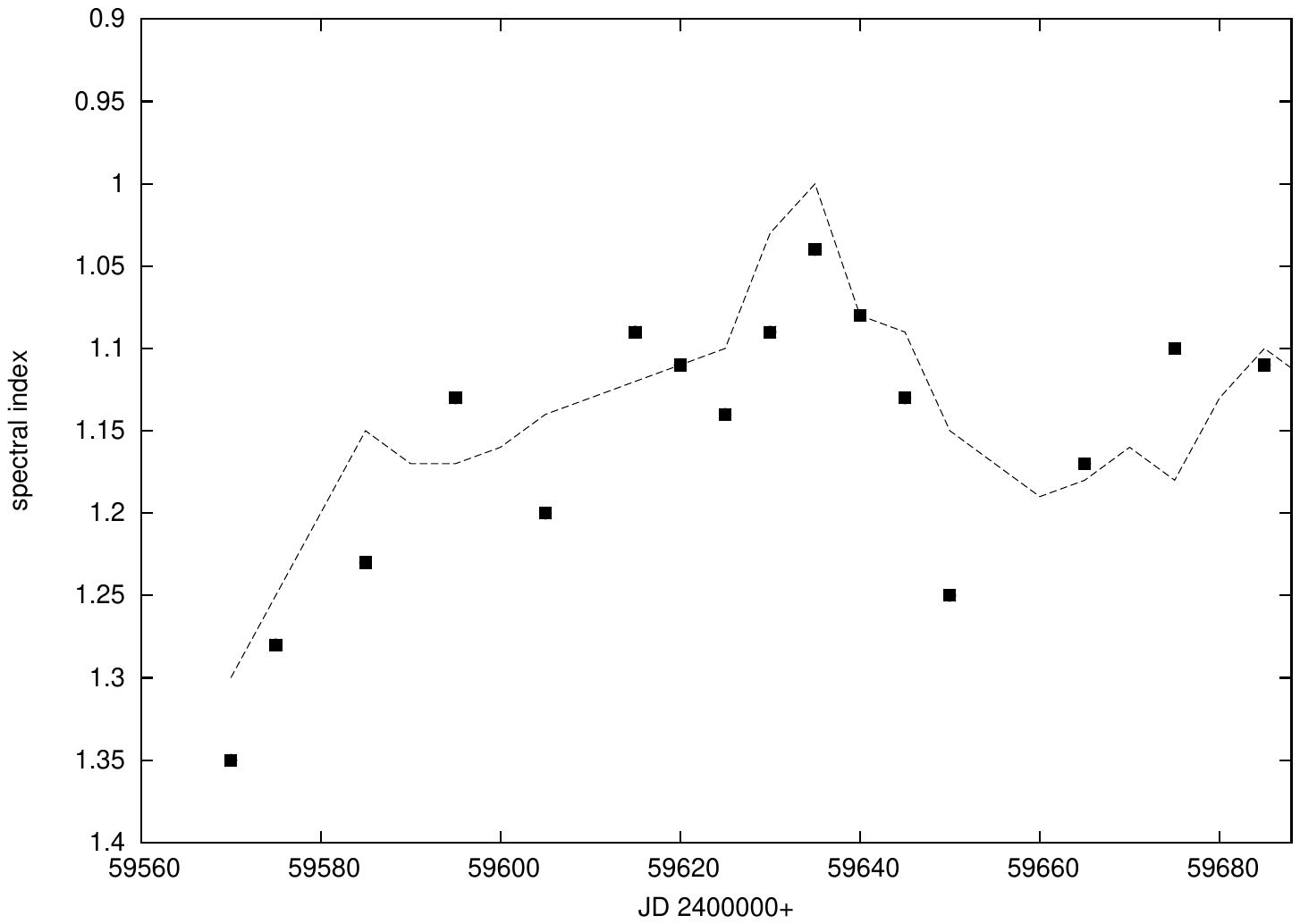}
\caption{The 
 calculated spectral index in 2022 based on a model where $\beta=0.7$ in the flare component and $\beta=1.35$ at the base level. The~line represents observations. The~errors are typically 0.03~units. \label{sp_2022}}
\end{figure}

We may then ask about the spectral index light curve in 2022 if the flux receives a contribution from the base level with $\beta=1.35$ and from the flare with the spectral index $\beta=0.7$, using the previously calculated flare vs. base split. We obtain the assumed flux of the flare from Figure~\ref{lc05_22} by drawing a straight line from (59570, 1.5) to (59655, 3.0), and~considering everything measured above this line to be the flare contribution. We obtain the points shown in Figure~\ref{sp_2022}. The~agreement with observations (line) is satisfactory. In~the data by \citet{val23}, there is one more brightness peak to the left of the range shown in Figure~\ref{sp_2022} at JD 59520. It has the spectral index $\beta=1.28$. There is no indication that other brightness peaks behave the same way as the JD 59635 peak. We also note that, from previous data of the spectra of OJ287, there is no evidence that the spectrum should turn bluer when the source is brighter in the 3--6 mJy range in the R-band \citep{Villforth2010}.

As we mentioned above, the~corresponding peaks between the 2005 and 2022 light curves should be separated by 6160 days. From~Figures~\ref{sp_2005} and \ref{sp_2022}, we see that this is indeed the case for the spectral index peaks, with an accuracy of one day.

\section{Expanding Cloud~Model}

Our disk models are based on magnetic accretion disks \citep{sak81,ste84,jia19,val19}. They are, in many ways, different from non-magnetic disks: much of the accretion power is channelled toward the upkeep of the hot corona, to~the jets and to the winds, while optical disk emission generally makes a minor contribution to the energy budget. Failure to realize this can lead to very wrong conclusions about the central engine of OJ287. Even reducing the central BH mass on this basis has been suggested \citep{Komossa2023}, but~the suggestion is totally unwarranted \citep{val23a}.

The blue flash arises when the radiating plasma has just burst out of the disk. This happens 0.11 yr after the disk impact in both 2005 and in~2022.

\begin{figure}[H]
\centering
\includegraphics[width=0.55\textwidth]{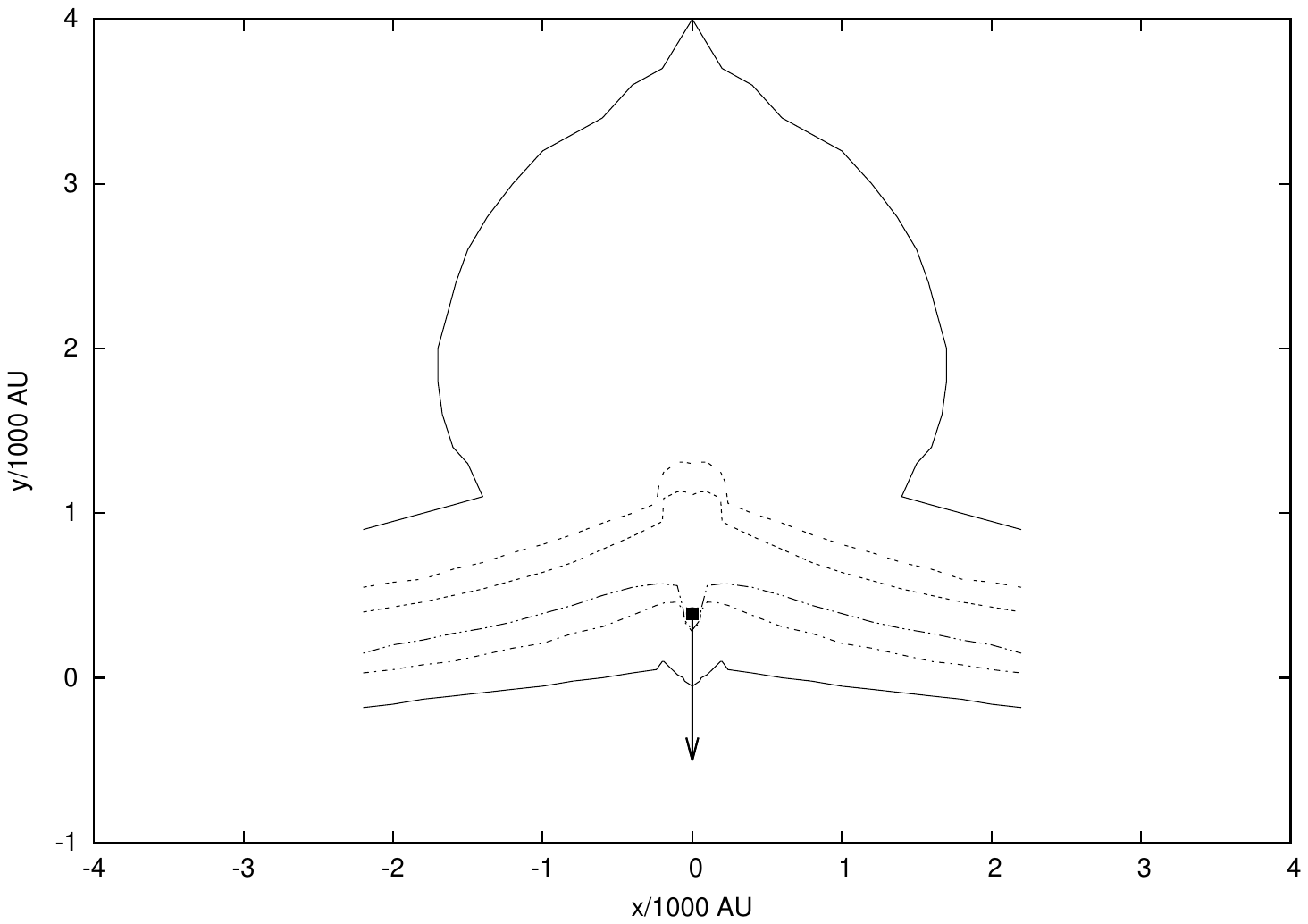}
\caption{Simulation 
 of the response of the accretion disk to the crossing of the secondary BH through the disk by \citet{iva98}. The~BH comes from above and crosses the disk vertically (see the arrow at the BH). The~disk is represented by density contours. The~outermost contour (solid line) is at the level of $10^{-6} \rho_{0}$, where $\rho_{0}$ is the density at the central plane. The~contours at the level of $0.125 \rho_{0}$ and $0.8 \rho_{0}$ (dotted and dash--dot lines) are also shown. The~density of the bubble is typically $4\times10^{-5} \rho_{0}$. The~central number density in our disk model is $\sim 2\times 10^{14} {\rm cm}^{-3}$ \cite{val19}.\label{bubble}}
\end{figure}

\citet{iva98} carried out hydrodynamical disk-impact simulations that are applicable to the OJ~287 situation \citep{val19}. In~Figure~\ref{bubble}, we illustrate the stage of the impact when the secondary has just come through the disk. Even though the impact is perpendicular to the disk, the model calculation should give an idea of what happens to the shocked gas in a slanted impact as well. Note that our viewpoint of the impact is from above, while the BH moves downwards in the~figure.

There is a rapidly expanding cloud of tenuous gas, which expands towards us~\cite{iva98}. In~our observation campaign, we set out to look for a signal from this thin high-velocity cloud. Since the disk is magnetic \citep{val19} and strong shocks accelerate particles, we expect to see a synchrotron source. More concretely, the~magnetic flux density $B$ at the center of the disk should be $B\sim10^7$ Gauss \citep{ste84,jia19}, and~if $B\sim\rho^{2/3}$, where $\rho$ is the density, we expect $B\sim10^4$ Gauss in the cloud. Compared with the radiating clouds in 3C273, where the magnetic flux density is $B\sim1$ Gauss \citep{sav08}, we can see that the peak emission moves from the radio frequencies in 3C273 to optical in OJ287, taking the peak frequency to be proportional to $B$.

Later during the passage of the BH through the accretion disk, a~gas bubble appears on the other side of the accretion disk, which can also provide a response in radiation. This is behind the disk from our direction of viewing, and~since the disk is optically thick, the~other cloud is not visible to~us.

\begin{figure}[H]
\centering
\includegraphics[width=0.65\textwidth]{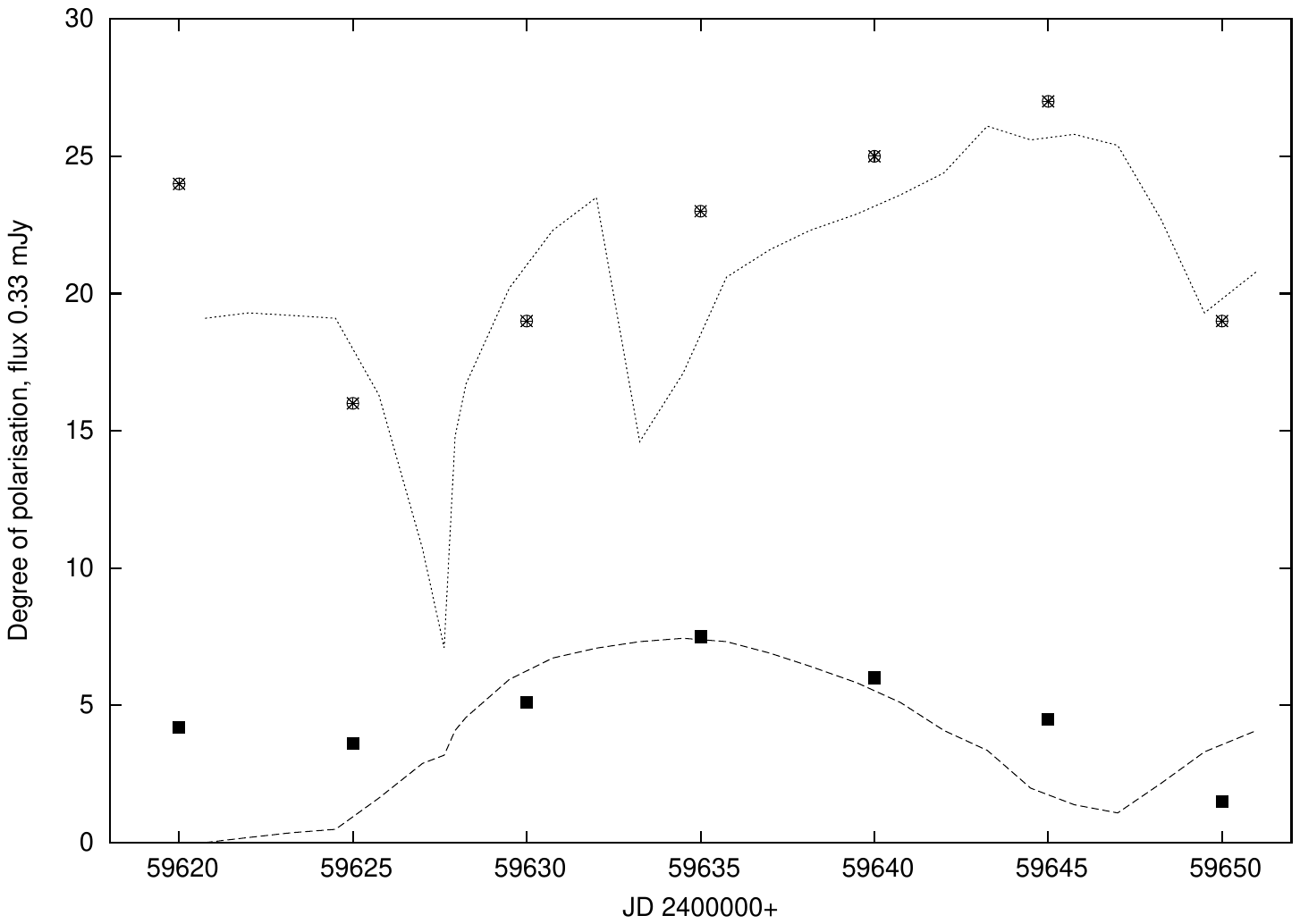}
\caption{A comparison 
 of the \citet{van71} expanding cloud model (lines) with observations (points) in 2022. The~lower line and the points refer to the R-band flux (remainder after subtracting the base level) while the upper line and the points present the degree of polarisation in percentages. The~errors in the latter points are typically two percentage points. Note that the degree of polarisation peaks later than the total flux both in the theory and observations. The~gap in observations between JD 59622 and JD 59632 prevents the verification of the polarisation minimum of the \citet{van71} model. The~observational points are averages in 5-day intervals from~\citet{val23}. The~flux values in the \citet{van71} model are arbitrarily~scaled.\label{vanderLaan}}
\end{figure}

We can see from Figure~\ref{bubble} that the low-density plasma spreads out in a rather semi-spherical manner in the space above the disk. The~radiation from such a process has been calculated by \citet{van71} for a radio flare in 3C273. In~3C273, we are discussing clouds of $\sim0.4$ pc in size \citep{sav08}; in~OJ287, the cloud is $\sim25$ times more compact. Therefore, the timescales are also contracted in OJ287 with respect to the 3C273 model by a factor of 25 in Figure~\ref{vanderLaan}, where we compare our observations of polarisation and total flux in 2022 with the van der Laan model of~3C273.

The special feature of this model is that the degree of polarisation does not follow the evolution of the total flux in a simple way. Of~course, this could also be attributed to processes occurring in relativistic jets, for~example,~due to shockwaves \citep{rai21}. However, here the expansion speeds are much slower than in jets~generally.

 We find that the polarisation behaves as expected, even though we missed the polarisation minimum in our observations. The~total flux model comes from the contribution of the presumably steady flux and the flare flux. By~adjusting the former contribution, one could bring down the total flux points closer to the zero level on the left-hand side of the~figure.

\section{Roche Lobe~Flares}

The primary jet is well observed from radio frequencies to X-rays, and extends to the megaparsec scales \citep{Marscher2011}. In~radio VLBI, as well as in optical polarisation, the jet is seen to wobble in a manner that can be explained by the influence of the secondary BH on the inner disk of the primary \citep{val12b,val13,Dey21}. The~jet points almost directly towards us, which means that occasionally the jet passes through our line of sight, resulting in a huge jump in the projected direction of the jet in the sky \citep{Agu12,Dey21}. These jumps take place at different times at different frequencies, which is understandable if the jet is helical and radiation at different frequencies arises at different parts of the helix \citep{val13}.

It has been suggested that the secondary BH also has a jet of its own \citep{vil98,Pih13,Pih13a}. While the spin of the primary is rather slow \citep{val16}, and~consequently the jet is somewhat weak in relation to its mass, at least in some jet models, we have good reasons to expect that the secondary black hole spins near its maximum speed, and~is therefore quite~bright.

The jet luminosity may be calculated in the \textit{Blandford-Znajek} process from the following expression \citep{gho97}:
\begin{equation}
L_{j} \sim m^{1.1} \dot{m}^{0.8} J^2                 
\end{equation}
where $m$ is the mass of the black hole, $\dot{m}$ the mass accretion rate and $J$ is its normalized spin, where $J = 0.38$ for the primary and $J \sim 1$ for the secondary (in the normalization, we divide by $m$, which necessarily takes a $J$ close to 1 or the maximum value, for~small $m$).
Using $\dot{m} = 1$ for the secondary (the Eddington rate) and $\dot{m} = 0.08$ for the primary \citep{val19} makes the secondary jet $43\%$ of the total~luminosity.

Thus, the total brightness of the secondary could be only some tens of percents of the primary's brightness. 
 Therefore, its influence on the total flux could be limited to certain orbital phases. One of the special phases is definitely the disk-crossing time. Then, the secondary has the possibility of accreting large amounts of gas from the primary disk at a rapid rate, leading to a large increase in brightness. On~the other hand, the~impact on the disk tends to strip the secondary disk from its outer layers, and~the disk-crossing related events should be~short-lived.

The observability of the secondary jet may also depend on its motion across the line of sight. In~the orbit model, the secondary exhibits typically a large transverse motion relative to line of sight, of~the order of 0.1 c, where c is the speed of light \citep{Dey18}. This results in the aberration of the jet of the order 0.1~radians. 

The Doppler factor of the primary jet has been estimated at $D = 17$, and~the viewing angle should be $\theta \sim 3.3$ degrees \citep{hov09}. This could be of the order of the beam width in the primary jet of OJ287. The~corresponding values for the secondary jet are not known. If~the secondary jet also has a narrow beam, this means that we are only able to see the full power of the secondary jet at certain phases of the~orbit. 

The viewing angle of the system is known from radio jet simulations: we know that the viewing direction is in the orbital plane about 0.1 radians forward from the axis of the primary disk \citep{Dey21}. Therefore, the secondary has to move in the same direction so that our line of sight coincides with the jet. Most recently, this happened in 2022 at the time of the disk-crossing. Before this time, we have to go back about half-a-century before we come to the previous time at which the secondary jet was seen head-on during a disk-crossing. Here, we assume that the secondary jet is lined up with the spin axis of the primary disk, which is the source of the accreting gas in the secondary \citep{Dey21}. The~recent RadioAstron map of OJ287 \citep{Gom22} may be seen as supporting this assumption \citep{val21a}.

\citet{Pih13} discussed the situation where the Roche lobe of the secondary BH is suddenly filled with plenty of gas, which is then accreted by the secondary in the sound-crossing time scale. This situation happened at 2021.78 when the secondary BH was 1000 AU above the disk and approaching it. A~rough estimate of the radius of the Roche lobe at that time is 1260 AU. In~the numerical simulation,~\citet{Pih13} found the accretion peak 0.08 yr later, which, in the present case, places the accretion peak at 2021.86. They estimate that the secondary could suddenly brighten up as a result of this new accretion and overpower the primary jet emission for a short period of time. The~time scales of the processes scale with the period of the innermost stable circular orbit (ISCO). 
For~the secondary BH, the~ISCO period is taken as 3.8 h, and~the accretion is expected to take place in less than 10 periods. Therefore, we expect events in time scales of a day or so, leading to what is usually referred to as intra-day variability (IDV).

Figure~\ref{lc05_22} shows a flare like this at JD 59531 or~at 2021.86. This is the time when the Roche lobe flare should have been observed. This IDV flare of 4.2 mJy is an order of magnitude bigger than the IDV flares usually seen in OJ287 and,~if regarded as a continuation of the normal IDV distribution, would represent over $10 \sigma$ deviation \citep{zen17}. Thus, it is most likely that we witness a different process~here.

The ISCO period for the primary black hole is $\sim100$ days, which makes it hard to model an over-$100\%$ flux change in the jet in such a short time scale \citep{Pih13}. Typical examples of large primary jet flares such as the 2016 and 2017 flares operate over a 100-day time scale, while large optical flares not associated with jets have the typical time scale of a week \citep{Dey18}. From~the structural point of view, the 2016 and 2017 flares do not belong to the latter set, in~contrast to the suggestion by \citet{Komossa2023}.

We searched the optical light curve of OJ287 during the OJ94 campaign~(1993--1998,~\citep{Pursimo2000,Pur21}), in~the 2005--2010 campaign \citep{val11b} and in the recent campaign (2015--2022, lead by S.Z.), where very dense monitoring was carried out over the total period corresponding to approximately 10 years worth of data. We found that there is only one flare that is even remotely similar to the 2021.86 flare in terms of amplitude (3.3 mJy in R) and rapidity (1~day): the~1993 December flare \citep{kid93,kid95}. 
This is one of the precursor flares that is thought to arise from the secondary jet also, as~we will discuss~below.

Note that the time of the Roche-lobe flare is theoretically specified within an interval of $\pm0.01$ yr. The~accidental chance of detecting an exceptional flare at this time is less than $0.2\%$, since the probability of finding such a flare in the 10 yrs of monitoring data is less than~unity.

We also checked when the corresponding flares would have taken place during the last 50 years, and~found that OJ287 was never observed exactly on the night when they should have~occurred.

\section{Precursors}

The authors of \citet{Pih13} also discussed precursor flares that arise when the secondary BH is roughly 4000 AU above the disk and on~the way down to impact on the disk.  Their cause is unknown, and since only three had been observed previously, it was unclear whether they are truly associated with the secondary. In~this respect, the prediction of the next precursor was significant. It was expected at the very end of~2020.

\begin{figure}
\centering
\includegraphics[width=0.65\textwidth]{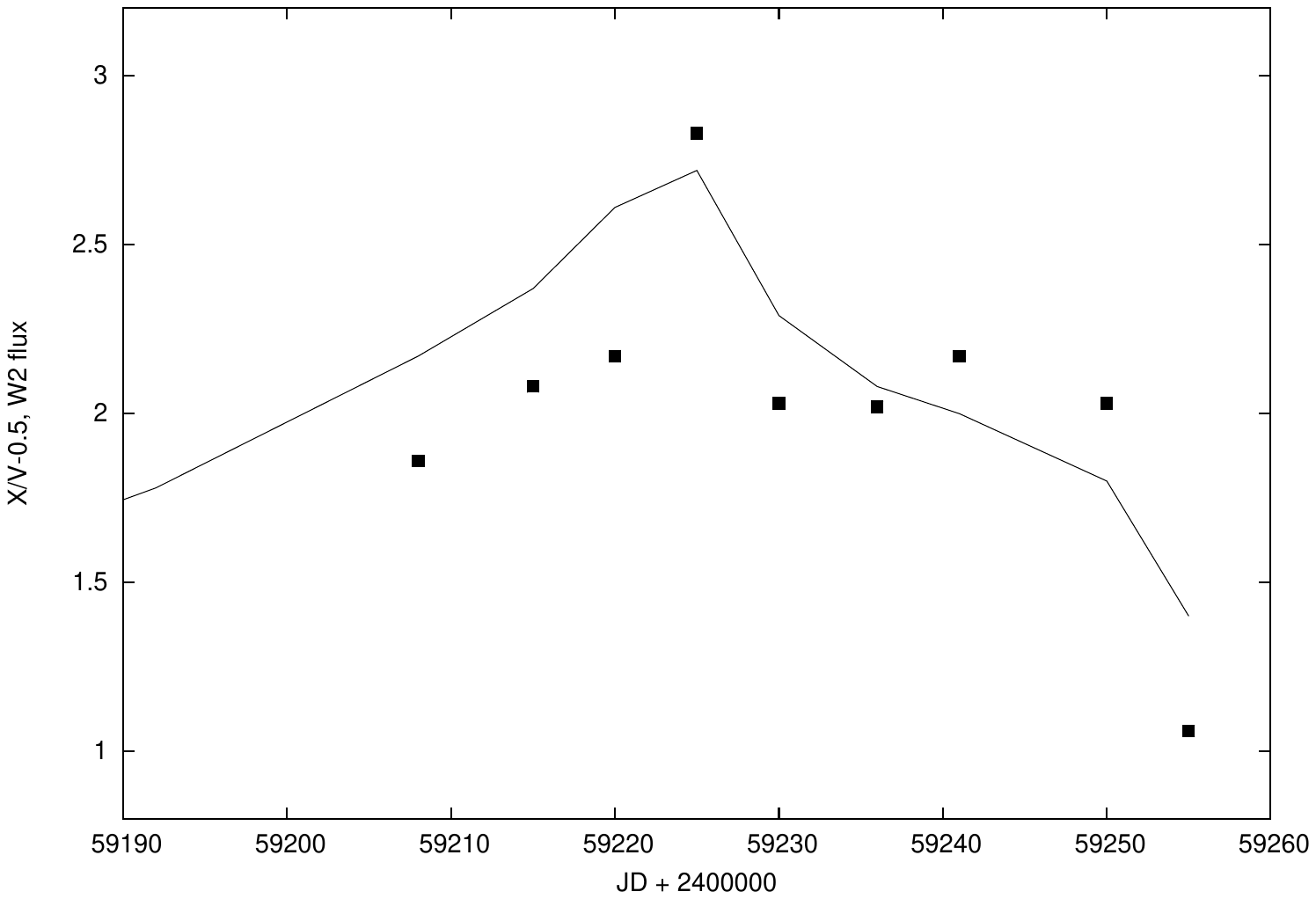}
\caption{The precursor 
 flare of 2021.02 in the UV band W2 (line) and in the V/X ratio from $\textit{Swift}$ observations (points, errors typically 0.3 units). The~simultaneous peaks in the two light curves suggests that the precursor flare emission is associated with the jet of the secondary black~hole.\label{precursor}}
\end{figure}

Figure~\ref{precursor} shows that the precursor flare may indeed have been observed. The~V/X ratio from $\textit{Swift}$ observations demonstrate that it is unlikely to be associated with the main jet. We may guess that there is some boundary at this level above the disk that activates the jet of the secondary when passing through~it.

\unskip

\section{Gamma-Ray~Flares}

\begin{figure}
\centering
\includegraphics[width=0.65\textwidth]{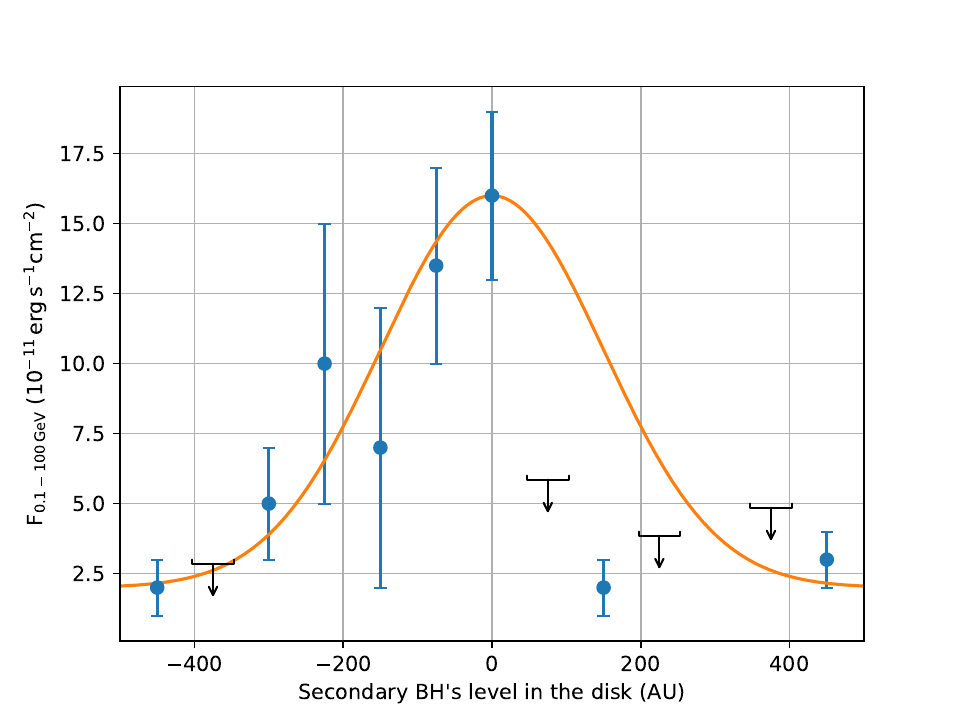}
\caption{Gamma-ray flux associated with the 2022.04 disk crossing. The~time axis was converted to the disk level height of the secondary according to the disk model of~\citet{val23}. The~line is the density profile of the disk in arbitrary~units.}
\label{oj287gamma}
\end{figure}

OJ~287 shows repeated flaring in the $\textit{Fermi}$ gamma-ray band \citep{Abdo2009, Agu11, Hodgson2017}, along with 
more quiescent epochs not detected by $\textit{Fermi}$. In~January 2022, $\textit{Fermi}$ observed the biggest flare in six years, and,~interestingly, this coincided with the expected disk impact at 2022.04 \citep{Komossa2022}. (This statement slightly depends on the binning of data that is used. Here, we are interested in long-lasting gamma-ray flares.)
Figure \ref {oj287gamma} shows the evolution of the flare. The~time axis was  converted to the disk level axis, showing where the secondary BH was at each stage of the flare. The~line shows the corresponding density profile of the~disk.

There is an obvious association between the disk density where the secondary BH is travelling and the strength of the gamma rays.
The important fact to notice here is that gamma-ray flares can be used as markers of the BH along its orbit. The~previous similar gamma-ray flare occurred during the July 2013 disk impact. The~\textit{Fermi} 7-day archives show gamma-ray flares on the weeks starting 4 July, 11 July and 18 July, with the~last one being the biggest (the biggest in 1.5 years). The~peak then corresponds to the date 2013.56. Even though these flares were not the biggest ever seen in OJ~287, they were, in some ways,~unique.

Looking at the $\textit{Fermi}$ 7-day archives, and~picking only those flares where the source was active on at least two successive weeks,  then the 2013 and 2022 flares are both among the six biggest flares, and~about equal in terms of total gamma-ray flux. The~main difference between them is that the 2013 flare covers 3 consecutive weeks while the 2022 flare covers only 2 weeks. Both follow the same density profile as shown in Figure~\ref{oj287gamma}, which is understandable since the speed of disk-crossing in 2022 was 1.5 times higher than the speed in 2013, according to Kepler's second~law. 

The gamma-ray flare coincides with a rather small optical flare. However, it is prominent in the polarisation light curve where the position angle undergoes a rapid swing toward the secondary jet at the position angle of 193 degrees~\cite{Dey21} and returns to normal equally quickly. At~the same time, the~degree of polarisation dropped dramatically, as~one would expect from the introduction of another component with a different angle of polarisation \citep{val23}. The~fall of the degree of polarisation during flares in OJ~287 is rare, although not~unique.

An obvious explanation of Figure~\ref{oj287gamma} is that the jet of the secondary BH meets increasingly denser material when it plunges into the disk. This matter is rotated into the jet stream, where collisions generate gamma rays. 
When the secondary reaches the central plane, the~jet is cut off due to the loss of the secondary disk in the collision. The~abrupt end of the gamma-ray flare has been seen in OJ287 on other occasions, but not among the brighter~flares.

For the peak gamma-ray flux, we may look at the results of~\citet{ara10}, who calculate the example of Centaurus A central source. In Centaurus A the BH mass was estimated to be similar to the mass of the secondary in OJ287. The gamma-ray luminosity measured by $\textit{Fermi}$ is $4\times10^{40}$ ergs sec$^{-1}$. \citet{ara10} use the cloud density 10$^{10}$ cm$^{-3}$ flowing into the jet with the speed 10$^9$ cm sec$^{-1}$.
In our case, the corresponding numbers are $2\times10^{14}$ cm$^{-3}$ and  $5\times10^9$ cm/sec \citep{val19}, increasing the jet-crossing mass flux by 10$^5$, and~presumably increasing the gamma-ray flux by a similar amount, to~$4\times10^{45}$ ergs sec$^{-1}$. This may be compared with the typical OJ287 gamma-ray flare luminosity of  $1.5\times10^{45}$ ergs sec$^{-1}$ \citep{sba12}.

Gamma-rays may be generated in OJ287 by many different mechanisms, which are usually associated with gas clouds or even stars entering the primary jet \citep{ban16,Hodgson2017,bed21,abd21}, and they may have different light curve shapes \citep{sai13,sai15}. This may also be the case at JD 59600, but~how likely is it that a gamma-ray flare would arise in OJ287 just at this time? Flares of this magnitude or bigger occurred in OJ287 six times during 2008--2022 ($Fermi$ archive, 7-day averages), including the 2013 and 2022 flares;~the background rate is about 1 per 1000 days.  The~accuracy of the orbit is about 3 days. This makes the probability of the accidental occurrence of a gamma-ray flare in the primary jet at exactly this time about 1 chance in 300, i.e.,~the coincidence is about 3 sigma effect for~each gamma-ray event. Taken together, the~probability of a chance coincidence is much~smaller.

\section{Past and Future Light Curve~Events}

In Table~\ref{tab1}, we list all impacts between 1970 and 2050. The~first column is the flare name and second column is the impact time in fractional years. In~the third column, we present the times of the precursor flares \citep{Pih13}. They may be associated with the secondary crossing of a magnetospheric boundary layer, but~their origin is not entirely clear. The~latest precursor took place before our campaign and~showed up as a prominent UV flare, as~we mentioned~above.


Column four lists the times of Roche lobe flares, with~brackets indicating a flare taking place behind the screen of the primary disk. Column five gives the times of the blue flashes, and~column six gives the time of the beginning of a major bremsstrahlung flare \citep{Dey18}. The~other flares (columns 3--5) are not as bright in  R-band flux as the major bremsstrahlung flare, but~they stand out in other ways. The~tidal flares are thought to arise from increased accretion flow, instigated by the influence of the secondary BH. They are listed in column 7 and can be as bright as the associated major flares, but~their time scales are much longer and they are easily differentiated from major flares by other observational~properties.

A good example is the pair of tidal flares in 2016 and 2017, which are so different from most big flares in the historical light curve that there is no way that one could put them in the same category, contrary to the claim by \citet{Komossa2023}. The~big flares in the historical record have a definite duration (about a week) and light curve shape \citep{Dey18}, and~they start by decreasing percentage polarisation, consistent with the flare component being unpolarised \citep{val08,val16}. In~contrast, the~tidal-flare time scales are about 3 months, and~their light curve profiles vary. The~degree of polarisation goes up in such flares \citep{val17}.

\begin{table}[H] 
\centering
\caption{Times of disk impacts, precursors, Roche-lobe flares, blue flashes, impact (bremsstrahlung) flares and tidal~flares\label{tab1}}
\begin{tabular*}{\hsize}{c@{\extracolsep{\fill}} cccccc}
\toprule
\textbf{Year}	& \textbf{Imp Time}  & \textbf{Precursor}   & \textbf{Roche-Lobe}  & \textbf{Blue Flash}  & \textbf{Imp Flare} & \textbf{Tidal Flare}\\
\midrule

1971  & 1971.13     & 1970.77  & 1971.10   & 1971.17   & 1971.13  & 1971.85\\
1973  & 1972.69   & 1971.77  & (1972.59) & 1972.77  & 1972.93  & 1974.5\\
1983  & 1982.96  & 1982.53  & 1982.92 & 1983.00  & 1982.96   & 1983.9\\
1984  & 1984.06  &  ------      &   (1984.01) & 1984.11  & 1984.12   & 1985.1\\
1994  & 1994.53  & 1993.92  & 1994.46   & 1994.60  & 1994.59  & 1995.6\\
1995  & 1995.81  &  ------      &     (1995.78)  & 1995.85  & 1995.84  & 1996.6\\
2005  & 2005.17  & 2004.19   & 2004.98  & 2005.28   & 2005.74  & 2007.7\\
2007  & 2007.67  & 2007.28   & (2007.64)  & 2007.70  & 2007.69 & 2008.45\\
2015  & 2013.56  & 2012.20    & 2013.22   & 2013.62  & 2015.87   & 2017.2\\
2019  & 2019.55  & 2019.20   & (2019.52)  & 2019.58  & 2019.57  & 2020.3\\
2022  & 2022.04  & 2020.93  & 2021.86    & 2022.15   & 2022.52  & 2024.5\\
2031  & 2031.38  & 2031.01   & (2031.35)  & 2031.42   & 2031.41   & 2032.3\\
2032  & 2032.67  & ------          & 2032.60  & 2032.74   & 2032.73  & 2034.1\\
2043  & 2043.09 & 2042.62  & (2043.04) & 2043.14  & 2043.15   & 2044.1\\
2044  & 2044.19  & ------         & 2044.15  & 2044.23  & 2044.20  & 2045.1\\
\bottomrule
\end{tabular*}
\end{table}

\section{Discussion and~Conclusions}

The 2021/2022 observing campaign was a great success and firmly confirmed the orbit model of~\citet{Dey18}. The~detection of the signals from the actual disk-crossing was important, since previously only delayed responses from the impacts have been reported. Both the 18 July 2013 and 19 January 2022 gamma-ray events agree, within an acceptable margin of error, with these two disk-crossing times. 
 Confirming the former flare is especially important since it leads directly to the determination of the spin of the primary black hole \citep{val16}.

The detection of the blue flashes was also important, since it tied together the orbital phases in 2005 and in 2022. We may regard the 2005 impact as well studied. This allows for us to determine the time of the secondary BH impact 
on the accretion disk in 2022. The~result is that the blue flash marks the emergence of hot radiating plasma from the accretion disk. This picture is well connected to the theoretical calculations of~\citet{van71} and \citet{iva98}.

The big surprise in this campaign was the large one-day flare on 11 November, 2021. The authors of \citet{Pih13} had previously discussed the possibility of such flares, but~when it happened, it was still unexpected. After~all, OJ~287 has now been closely followed for over 50 years and one would not expect new phenomena to turn up, especially in optical photometry, which has been the most~complete area of study. 

The explanation for this is the required orbital phase: in order to brighten up the secondary, the~BH must suddenly obtain a huge amount of gas for accretion. This happens once per decade on the near side. Whether a similar brightening can also be seen through the accretion disk screen between us and the secondary is an interesting question. Table~\ref{tab1} lists the times at which a one-day flares might have appeared in the~past.

For the tidal flares, the~predicted light curve has maxima at 2023.5 and 2024.2 \citep{Pih13}. However, the~tidal flare predictions are not very accurate. These would be the last remnants of the latest activity cycle. Then, OJ~287 should return to new activity again during the 2031~cycle.

\vspace{6pt} 



\authorcontributions{M.J.V. was responsible for the main design of this article, while L.D. was in charge of orbit calculations with advice from A.G.; S.Z coordinated the R-band observing campaign, with participants T.P., E.K., D.E.R., V.V.K., K.M., M.D., M.M., A.S. and M.Z.; A.L., M.T. and J.L.G. were responsible for radio observations, while A.C.G, A.V.B., R.I. and M.U. carried out polarimetric observations. The historical data base, including the 2005 light curve, come from R.H., S.C., A.S., H.J.L. and K.N.: M.J. and J.\v{S} provided the BVRI photometry. Several authors contributed to theoretical~ideas. All authors have read and agreed to the published version of the manuscript.} 

\funding{
L.D. and A.G. acknowledge the support of the Department of Atomic Energy, Government of India, under~project identification \# RTI 4002.
A.G. is grateful for the financial support and hospitality of the Pauli Center for Theoretical Studies and the University of Zurich.
S.Z. acknowledge the grant NCN 2018/29/B/ST9/01793.
R.H. acknowledges support from the European Union’s Horizon 2020 Programme under the AHEAD2020 project (grant agreement n. 871158).
}

\dataavailability{Data is available from authors upon a reasonable request. 
} 

\acknowledgments{We thank Helen Jermak and Callum McCall for providing polarisation data during the campaign period. We also thank Manpreet Singh for help with analysing \textit{Swift} data.}

\conflictsofinterest{Alberto Sadun is on the editorial board of Galaxies, and~one of the guest editors for this Special~Issue.}

\reftitle{References}

\end{paracol}
\end{document}